\documentclass[sigconf,nonacm]{acmart}

\usepackage{amsmath,amsfonts}
\usepackage{algorithmic}
\usepackage{graphicx}
\usepackage{textcomp}
\usepackage{xcolor}
\usepackage{weiwAlgorithm}
\usepackage{subfigure}
\usepackage{listings}
\usepackage{enumitem}
\usepackage{amsthm}

\AtBeginDocument{%
  }

\setcopyright{acmlicensed}
\copyrightyear{2018}
\acmYear{2018}
\acmDOI{XXXXXXX.XXXXXXX}

\acmConference[Sigmod '25]{International Conference on Management of Data}{June 22--27,
  2025}{Berlin, Germany}
\setlist[itemize]{leftmargin=1.2em, labelsep=0.5em}

\lstdefinelanguage{gopt}
{
  morekeywords={
    getV,
    expandE,
    patternStart,
    patternEnd,
    UNION,
    MATCH,
    RETURN,
    WHERE,
    AND,
    OR,
    SET,
    MERGE,
    ON,
    IN,
    class,
    implements,
    extends,
    interface,
    return,
    override,
    Double,
    join,
    select,
    group,
    order,
    new
  },
  morecomment=[l]{//},
}

\definecolor{eclipseBlue}{RGB}{42,0.0,255}
\definecolor{eclipseGreen}{RGB}{63,127,95}
\definecolor{eclipsePurple}{RGB}{127,0,85}

\begin{document}

\title{A Modular Graph-Native Query Optimization Framework}

\author{Bingqing Lyu, Xiaoli Zhou, Longbin Lai, Yufan Yang, Yunkai Lou, Wenyuan Yu, Ying Zhang$^{\ddag}$, Jingren Zhou}
\email{{bingqing.lbq,yihe.zxl, longbin.lailb, xiaofan.yyf, louyunkai.lyk, wenyuan.ywy, jingren.zhou}@alibaba-inc.com}
\email{ying.zhang@zjgsu.edu.cn}
\affiliation{%
  \institution{Alibaba Group, $^{\ddag}$Zhejiang Gongshang University}
  \country{China}
}


%%%%%%%%%%%%%%%%%%%%%%%%%%%%%%%%%%%%%%%%%%%%%%%%%%%%%%%%%%%%%%%%%%%%%%%%
% COMMENTS
%%%%%%%%%%%%%%%%%%%%%%%%%%%%%%%%%%%%%%%%%%%%%%%%%%%%%%%%%%%%%%%%%%%%%%%%

% \settopmatter{printacmref=false}
% \setcopyright{none}

\begin{abstract}
  \underline{C}omplex \underline{G}raph \underline{P}atterns (\cgps), which combine pattern matching with relational operations, are widely used in real-world applications.
  % These patterns enable users to identify arbitrary patterns in a graph and further perform in-depth relational analysis on the results.
  Existing systems rely on monolithic architectures for \cgps, which restrict their ability to integrate multiple query languages and lack certain advanced optimization techniques.
  Therefore, to address these issues, we introduce \gopt, a modular graph-native query optimization framework with the following features:
  (1) support for queries in multiple query languages, (2) decoupling execution from specific graph systems, and (3) integration of advanced optimization techniques.
  Specifically, \gopt~offers a high-level interface, \graphirbuilder, for converting queries from various graph query languages into a unified intermediate representation (\ir), thereby streamlining the optimization process.
  It also provides a low-level interface, \physicalbuilder, enabling backends to register backend-specific physical operators and cost models.
  Moreover, \gopt~employs a graph-native optimizer that encompasses extensive heuristic rules, an automatic type inference approach, and cost-based optimization techniques tailored for \cgps.
  Comprehensive experiments show that integrating \gopt~significantly boosts performance, with Neo4j achieving an average speedup of $9.2\times$ (up to $48.6\times$), and \gs~achieving an average speedup of $33.4\times$ (up to $78.7\times$), on real-world datasets.
\end{abstract}

%TODO: generate CCSXML

%   <ccs2012>
%      <concept>
%          <concept_id>10002951.10002952.10003190.10003192.10003210</concept_id>
%          <concept_desc>Information systems~Query optimization</concept_desc>
%          <concept_significance>500</concept_significance>
%          </concept>
%    </ccs2012>
% \ccsdesc[500]{Information systems~Query optimization}

%\keywords{Graph Query Optimization, Graph Pattern Matching, Graph Database}


\newcommand{\stitle}[1]{\vspace{0.5ex}\noindent{\bf #1}}
\newcommand{\etitle}[1]{\vspace{0.5ex}\noindent{\em\underline{#1}}}
\newcommand{\eetitle}[1]{\vspace{0.5ex}\noindent{\em{#1}}}
\newcommand{\eat}[1]{}

\newcommand{\patrel}{{\kw{CGP}}}
\newcommand{\patrels}{{\kw{CGPs}}}
\newcommand{\bgp}{{\kw{pattern}}}
\newcommand{\bgps}{{\kw{patterns}}}
\newcommand{\cgp}{{\kw{CGP}}}
\newcommand{\cgps}{{\kw{CGPs}}}
\newcommand{\estimation}{{\kw{CardinalityEstimation}}}

\newcommand{\kw}[1]{\textsf{#1}\xspace}
\newcommand{\gs}{{\kw{GraphScope}}}
\newcommand{\gaia}{{\kw{Gaia}}}
\newcommand{\glogue}{\kw{GLogue}}
\newcommand{\gopt}{\kw{GOpt}}
\newcommand{\glogs}{\kw{GLogS}}
\newcommand{\cypherplanner}{\kw{CypherPlanner}}

\newcommand{\btype}{\kw{BasicType}}
\newcommand{\btypes}{\kw{BasicTypes}}
\newcommand{\utype}{\kw{UnionType}}
\newcommand{\utypes}{\kw{UnionTypes}}
\newcommand{\unionall}{\kw{AllType}}
\newcommand{\unionalls}{\kw{AllTypes}}
\newcommand{\xtype}{\kw{DynType}}
\newcommand{\elemtype}{\lambda}
\newcommand{\type}{\tau}
\newcommand{\prop}{\pi}

\newcommand{\graphtype}{\kw{T}}
\newcommand{\id}{\kw{ID}}

\newcommand{\nbr}[1]{\kw{$N_{#1}$}}
\newcommand{\nbre}[1]{\kw{$N_{#1}^E$}}
\newcommand{\adj}{\kw{Adj}}

\newcommand{\filterrule}{\kw{FilterIntoPattern}}
\newcommand{\fusionrule}{\kw{ExpandGetVFusion}}
\newcommand{\trimrule}{\kw{FieldTrim}}
\newcommand{\commonrule}{\kw{ComSubPattern}}
\newcommand{\selectpush}{\kw{FilterPushDown}}
\newcommand{\joinrule}{\kw{PatternJoin}}
\newcommand{\joinelimrule}{\kw{JoinToPattern}}

\newcommand{\freq}{\mathcal{F}}
\newcommand{\cost}{\kw{Cost}}

\newcommand{\getnbrtype}{\kw{getOutNbr}}
\newcommand{\getnbretype}{\kw{getOutEdge}}
\newcommand{\glogedge}{\kw{getEdges}}
\newcommand{\glogvertex}{\kw{getFreq}}
\newcommand{\getcount}{\kw{getFreq}}
\newcommand{\getcandi}{\kw{getCands}}
\newcommand{\getcost}{\kw{getCost}}
\newcommand{\computecost}{\kw{computeCost}}

\newcommand{\gloguequery}{\kw{GLogueQuery}}

\newcommand{\planmap}{\kw{m}}
\newcommand{\lbound}{\kw{getLowerBound}}
\newcommand{\candi}{\kw{CandVTypes}}
\newcommand{\candie}{\kw{CandETypes}}

\newcommand{\init}{\kw{GreedyInitial}}

\newcommand{\code}[1]{\texttt{#1}}
\newcommand{\traits}{{\code{traits}}}
\newcommand{\trait}{{\code{trait}}}
\newcommand{\graphirbuilder}{\code{GraphIrBuilder}}
\newcommand{\physicalbuilder}{\code{PhysicalSpec}}
\newcommand{\physicalspec}{\code{PhysicalSpec}}
\newcommand{\ir}{\kw{GIR}}

\newcommand{\wcoj}{\kw{WcoJoin}}

% logical operators
\newcommand{\kk}[1]{\texttt{#1}}
\newcommand{\scan}{{\kk{SCAN}}}
\newcommand{\expandedge}{{\kk{EXPAND}\_\kk{EDGE}}}
\newcommand{\getvertex}{{\kk{GET}\_\kk{VERTEX}}}
\newcommand{\expandvertex}{{\kk{EXPAND}}}
\newcommand{\expandpath}{{\kk{EXPAND}\_\kk{PATH}}}
\newcommand{\matchpattern}{{\kk{MATCH}\_\kk{PATTERN}}}
\newcommand{\pattern}{{\kk{BGP}}}
\newcommand{\matchstart}{{\kk{MATCH}\_\kk{START}}}
\newcommand{\matchend}{{\kk{MATCH}\_\kk{END}}}
\newcommand{\pathstart}{{\kk{PATH}\_\kk{START}}}
\newcommand{\pathend}{{\kk{PATH}\_\kk{END}}}
\newcommand{\project}{\kk{PROJECT}}
\newcommand{\select}{\kk{SELECT}}
\newcommand{\order}{\kk{ORDER}}
\newcommand{\limit}{\kk{LIMIT}}
\newcommand{\optional}{\kk{OPTIONAL}}
\newcommand{\difference}{\kk{DIFFERENCE}}
\newcommand{\group}{\kk{GROUP}}
\newcommand{\joinopr}{\kk{JOIN}}
\newcommand{\union}{\kk{UNION}}
\newcommand{\map}{\kk{MAP}}
\newcommand{\flatmap}{\kk{FLATMAP}}
\newcommand{\source}{\kk{SOURCE}}

% physical operators
\newcommand{\expandinto}{\kw{ExpandInto}}
\newcommand{\expandintersect}{\kw{ExpandIntersect}}
\newcommand{\physicalexpand}{\kw{Expand}}
\newcommand{\hashjoin}{\kw{HashJoin}}
\newcommand{\physicalselect}{\kw{Select}}
\newcommand{\join}{\kw{Join}}
\newcommand{\expand}{\kw{Expand}}

\newcommand{\goptplan}{\kw{GOpt-plan}}
\newcommand{\gsplan}{\kw{GS-plan}}
\newcommand{\neoplan}{\kw{Neo4j-plan}}

\newcommand{\reffig}[1]{Fig.~\ref{fig:#1}}
\newcommand{\refsec}[1]{Section~\ref{sec:#1}}
\newcommand{\reftab}[1]{Table~\ref{tab:#1}}
\newcommand{\refalg}[1]{Algorithm~\ref{alg:#1}}
\newcommand{\refeq}[1]{Eq.~\ref{eq:#1}}
\newcommand{\refdef}[1]{Definiton~\ref{def:#1}}
\newcommand{\refthm}[1]{Theorem~\ref{thm:#1}}
\newcommand{\reflem}[1]{Lemma~\ref{lem:#1}}
\newcommand{\refex}[1]{Example~\ref{ex:#1}}
\newcommand{\refpro}[1]{Property~\ref{pro:#1}}
\newcommand{\refrem}[1]{Remark~\ref{rem:#1}}

\newcommand{\ie}{\emph{i.e.,}\xspace}
\newcommand{\eg}{\emph{e.g.,}\xspace}
\newcommand{\wrt}{\emph{w.r.t.}\xspace}
\newcommand{\aka}{\emph{a.k.a.}\xspace}
\newcommand{\kwlog}{\emph{w.l.o.g.}\xspace}
\newcommand{\kwhp}{\emph{w.h.p.}\xspace}

\newcommand{\todo}[1]{\textcolor{red}{$\Rightarrow$#1}}
\newcommand{\tbf}{\textbf{\textcolor{red}{X}}\xspace}
\newcommand{\warn}[1]{{\color{red}{#1}}}
\newcommand{\revise}[1]{{\textcolor{blue}{#1}}}
\newcommand{\lourevise}[1]{{\textcolor{orange}{#1}}}

\newcommand{\oom}{\texttt{OOM}\xspace}
\newcommand{\ot}{\texttt{OT}\xspace}
\newcommand{\rbo}{\kw{RBO}\xspace}
\newcommand{\cbo}{\kw{CBO}\xspace}

\newcommand{\hybrid}{\emph{Hybrid Semantics}\xspace}
\newcommand{\arbitrary}{\emph{Arbitrary Types}\xspace}

\newcommand{\vtype}[1]{\textit{#1}\xspace}
\newcommand{\company}{Alibaba}

\newcommand{\gsgopt}{\kw{GraphScope-GOpt}}
\newcommand{\neogopt}{\kw{Neo4j-GOpt}}

\newtheorem{remark}{Remark}[section]

\maketitle

\section{Introduction}
\label{sec:intro}

Graph databases~\cite{neo4j,tigergraph,janusgraph,neptune,graphflow} have emerged as powerful tools for analyzing complex relationships across various domains.
At the core of these databases lies graph pattern matching, a fundamental and extensively researched problem~\cite{lai2019distributed,aberger2017emptyheaded,ullmann1976algorithm,shang2008quicksi,yang2021huge,glogs,han13turbo,ammar2018distributed,meng2024survey}.
{Graph \bgp~ (abbr.~\bgp~ henceforth)}
matching aims to identify all subgraphs in a data graph that match a given query pattern and is widely used in numerous areas~\cite{flake02www,hu18local,DBLP:journals/csur/CannataroGV10,prvzulj2006efficient,DBLP:journals/jcisd/ZhuYYLCCL05,gauzre15}.

\begin{figure}[t]
  \centering
  \includegraphics[width=0.85\linewidth]{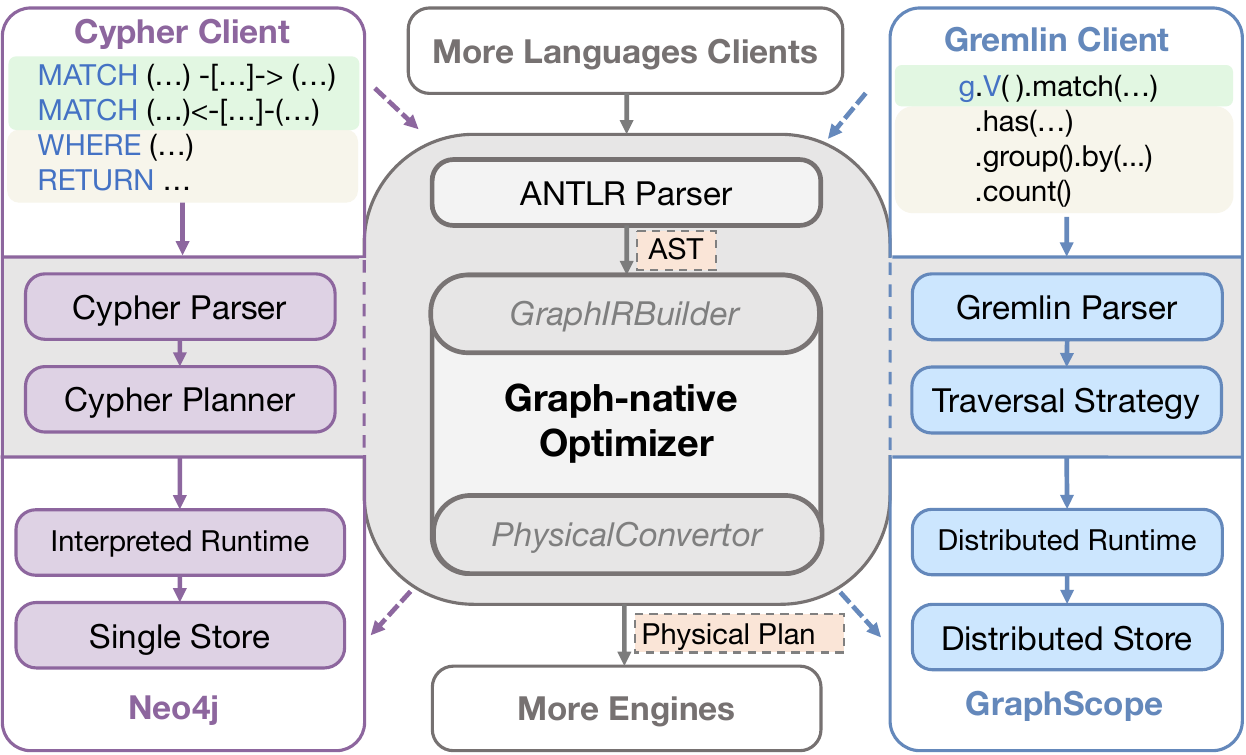}
  % \vspace*{-4.5mm}
  \caption{An example of processing \cgps~in the monolithic systems of Neo4j and \gs, compared with the modular design of \gopt.}
  \vspace*{-3mm}
  \label{fig:example}
\end{figure}

Recently, there has been growing interest in complex graph patterns (\cgps) due to their utility in real-world applications~\cite{angles17,pgql,gcore,gql,gsql,gremlin}.
\cgps~ build upon {patterns} by incorporating additional relational operations such as projection, selection, and aggregation, enabling more expressive and versatile query capabilities.
To enable efficient expression and execution of \cgps, existing graph databases typically adopt a \emph{monolithic} architecture design, where the entire process is tightly integrated from query parsing to execution.
These systems handle specific query languages, starting with a parser that translates the query into an internal representation.
A query optimizer then generates a physical execution plan, which is subsequently executed on an underlying execution engine.
Two representative examples of graph systems are Neo4j~\cite{neo4j} and \gs~\cite{graphscope}, illustrated in \reffig{example}.

Neo4j processes \cgps~ written in the Cypher query language\cite{cypher}.
It employs a Cypher parser to convert the query into an intermediate structure, uses a \cypherplanner to optimize queries through heuristic rules and cost-based techniques, and executes the optimized plans on its single-machine backend.
In contrast, \gs~\cite{graphscope} is designed for industrial-scale distributed graph processing.
It supports \cgps~ expressed in Gremlin~\cite{gremlin}, utilizes a Gremlin parser, applies \code{TraversalStrategy} to register heuristic optimization rules, and distributes the execution across its dataflow-based distributed backend engine~\cite{qian2021gaia}.

\stitle{Motivation and Challenges.}
The monolithic architecture used by existing systems like Neo4j and \gs~ for processing \cgps~ poses two key limitations.

\etitle{Limitation 1}.
In practice, while most graph systems are designed to support a single query language, there are scenarios where supporting multiple query languages becomes necessary to meet evolving user demands.
For example, at \company, users transitioning from Neo4j to \gs~ for scalable query processing may require support for Cypher queries, despite \gs~ being originally designed for Gremlin.
Additionally, the upcoming standardization of the graph query language GQL~\cite{gql} further drives the need for systems to adapt and expand their capabilities. {While a language converter is feasible, the only official converter currently available translates Cypher to Gremlin~\cite{cypher_for_gremlin}, limiting its applicability in other scenarios, such as adopting GQL in \gs.
The monolithic architectures make it challenging for existing systems to integrate support for additional query languages.}
%For instance, \gs~ is tightly integrated with a Gremlin parser and relies on a traversal-step-based intermediate representation.

\etitle{Limitation 2}.
Different systems employ distinct optimization strategies for \cgps, each with strengths, but often lack advanced techniques available in others.
For example, Neo4j leverages cost-based optimization {(CBO)} to determine the optimal search order but lacks advanced techniques like worst-case optimal join~\cite{ngo2018worst}, which \gs~ employs for efficient cyclic pattern matching.
Conversely, \gs~ relies on a rule-based optimizer, missing the sophisticated {CBO} strategies available in Neo4j.
Both systems also lack support for recent advancements such as high-order statistics~\cite{glogs}, critical for improving cardinality estimation and query planning.
The monolithic architectures make it difficult to integrate such optimization techniques seamlessly.

A potential solution is to adopt a \emph{modular} and \emph{unified} optimization framework with key features:
\begin{itemize}[noitemsep,topsep=0pt]
\item \textbf{Support for multiple query languages}: The framework should inherently support parsing and processing multiple query languages, enabling seamless interoperability.
\item \textbf{Decoupling from existing systems}: By being independent of specific graph systems, it can provide modular interfaces that simplify integration with various platforms.
\item \textbf{Advanced optimization capabilities}: The framework should integrate advanced optimization techniques from academia and industry into a unified system.
\end{itemize}

Beyond the efforts of system design, the implementation of such a framework poses immediate challenges.

\etitle{\textbf{C1}: How to support cross-language compatibility?}
\cgps~ can be expressed in diverse query languages such as Cypher and Gremlin, each with distinct syntaxes and semantics, making integration into a unified framework challenging.

%\etitle{\textbf{C2}: How to exploit diverse optimization strategies?}
%Different graph systems employ unique optimization strategies, and re-implementing advanced techniques across all systems is redundant and inefficient.
%Due to their tightly coupled designs, current systems do not provide interfaces to plug in optimization strategies from others.
%While converting graph operations into relational equivalents~\cite{orientdb,DBLP:conf/sigmod/TianXZPTSKAP20,DBLP:conf/sigmod/SunFSKHX15} enables the use of relational optimization techniques, this non-graph-native approach may miss optimization opportunities specific to graph pattern matching~\cite{glogs,yang2021huge}.

\etitle{\textbf{C2}: How to effectively optimize \cgps?}
Optimizing \cgps~ is inherently complex due to their hybrid semantics, combining \bgps~ with relational operations.
Existing graph optimization techniques~\cite{yang2021huge,mhedhbi2019optimizing} focus on \bgps~ and often overlook relational aspects. While converting \bgps~into relational equivalents~\cite{orientdb,DBLP:conf/sigmod/TianXZPTSKAP20,DBLP:conf/sigmod/SunFSKHX15} enables the use of relational optimization techniques, this \emph{non-graph-native} approach may miss optimization opportunities specific to graph pattern matching~\cite{glogs,yang2021huge}.
Another challenge is handling arbitrary type constraints. \cgps~ often include \bgps~ without explicit type constraints when users have not assigned specific types on certain vertices and edges and can, in theory, match any type in the data graph. In practice, however, these elements are implicitly restricted by the underlying data.
Current techniques~\cite{glogs} depend heavily on explicit type constraints, limiting their effectiveness in such cases.

\stitle{Our Solutions.}
To address the challenges, {We present \gopt, a modular \emph{graph-native} optimization framework for \cgps, as illustrated in \reffig{example}. It is designed to support multiple query languages and seamlessly integrate with existing graph systems.}

\etitle{\textbf{S1}: Modular System Design.}
\gopt~ enables cross-language compatibility with a unified graph intermediate representation (\ir) for \cgps, combining \bgps~ and relational operations.
{\gopt~provides a high-level interface, \graphirbuilder, that translates the Abstract Syntax Tree (AST) generated by tools like ANTLR \cite{antlr} from various query languages into a unified \ir, thereby decoupling query processing from specific languages.}
Additionally, \gopt~ introduces a low-level interface, \physicalbuilder, enabling graph systems to register their backend-specific operators.
This flexibility allows integration with various graph systems while maintaining compatibility with multiple query languages.
Currently, \gopt~supports both Cypher and Gremlin and has been integrated into Neo4j and \gs. Note that the standardized GQL is not yet supported, as its official grammar interpretation tool has not been finalized.

\etitle{\textbf{S2}: Graph-native Optimization for \cgps.}
Building on the unified \ir, \gopt~ enables graph-native optimization by addressing the interaction of \bgps~ and relational operations in \cgps.
A comprehensive set of heuristic rules has been designed for the rule-based optimization {(RBO)}, which are extensible and pluggable.
To handle \bgps~ with arbitrary types, we introduce an automatic type inference algorithm that deduces appropriate type constraints for pattern vertices and edges from graph data.
%This algorithm leverages the underlying graph data and does not rely on schema-strict contexts, making it applicable for schema-loose systems like Neo4j~\cite{Angles_2023}.
By inferring type constraints, we can fully exploit {CBO} techniques, such as those in~\cite{glogs}, to enhance cardinality estimation using high-order statistics.
Using the estimated cardinalities, backend-specific physical operators, and their cost models, we propose a top-down search framework with branch-and-bound strategies to optimize \cgps, determining the optimal execution plan by minimizing the estimated cost.
The unified framework allows advanced graph optimization techniques to be seamlessly integrated, benefiting all integrated graph systems.

{
Concerns naturally arise regarding the unified design of \gopt, particularly in: (1) the theoretical completeness of \ir~in expressing diverse query languages with their syntactic and semantic differences; and (2) the compatibility with different data models, from Neo4j's schema-loose model to \gs's schema-strict model~\cite{Angles_2023}. In this paper, we briefly remark on these concerns and refer readers to our open-source project for more details~\cite{gopt_github}.
}

\stitle{Contributions and organization}.
Below, we summarize our contributions and the organization of the paper.

% \noindent(1) We have introduced \gopt, the first graph-native optimization framework, as far as we know, for industrial-scale graph database systems. 
% It supports multiple query languages via a unified intermediate representation and an extensible optimization architecture.

% \noindent(2) We have engineered advanced optimization techniques, including a comprehensive set of heuristic rules, an automatic type inference algorithm, and a novel cardinality estimation method, integrated into a cost-based optimizer with a top-down search framework.

% \noindent(3) Implementation of \gopt~(\refsec{impl}): We implement \gopt~as an open-source project~\cite{gopt_github} on top of Apache Calcite~\cite{Begoli_2018}, supporting Gremlin and Cypher, and integration with graph systems Neo4j and \gs.

% \noindent(4) Experimental evaluation (\refsec{exp}): We conduct extensive experiments on real-world datasets, demonstrating the significant performance gains by integrating \gopt, achieving an average speedup of $9.2\times$ (up to $48.6\times$) on Neo4j and $33.4\times$ (up to $78.7\times$) on \gs.
% Furthermore, \gopt~ has been successfully deployed with \gs~ to handle real-world applications at \company, demonstrating its practicality and scalability.

\noindent(1) Introduction of \gopt (Sec.\ref{sec:arch} and Sec. \ref{sec:trans}): \gopt~stands as the first, as far as we know, graph-native optimization framework for industrial-scale graph systems, %enabling support for multiple query languages through a unified intermediate representation and an extensible optimization architecture.
supporting multiple query languages through a unified intermediate representation and an extensible optimization architecture.

\noindent(2) Advanced optimization techniques (Sec.\ref{sec:opt}): \gopt~ introduces a comprehensive set of heuristic rules, an automatic type inference algorithm, and a novel cardinality estimation method, integrated into a cost-based optimizer with a top-down search framework.

\noindent(3) Implementation and experiments (Sec. \ref{sec:impl} and Sec. \ref{sec:exp}): We implemented \gopt~as an open-source project~\cite{gopt_github} atop Apache Calcite~\cite{Begoli_2018}, supporting Gremlin and Cypher, and integrating with Neo4j and \gs. Experiments show significant performance gains, with \gopt~achieving an average speedup of $9.2\times$ (up to $48.6\times$) on Neo4j and $33.4\times$ (up to $78.7\times$) on \gs~using real-world datasets.
\gopt~ has also been deployed in real applications at \company, demonstrating its practicality and scalability.

Additionally, \refsec{background} covers background and related work, \refsec{preliminaries} introduces the notations and definitions used in this paper, and \refsec{con} concludes the paper.

% The paper is organized as follows: \refsec{background} covers background and related work, {\refsec{preliminaries} introduces the notations and definitions used in this paper}, \refsec{arch} provides an overview of the system architecture, \refsec{trans} describes intermediate representation and query transformation, \refsec{opt} details optimization techniques, \refsec{impl} discusses system implementation, \refsec{exp} presents experimental results,
% % \refsec{rel} reviews related works,
% and \refsec{con} concludes the paper.

\section{Background and Related Work}
\label{sec:background}
Corresponding to the main challenges in implementing a modular and unified optimization framework mentioned above, this section reviews the existing query languages and optimization techniques, then presents the existing databases for optimizing \cgps.

% As stated in \refsec{intro}, the main challenges in implementing a modular and unified optimization framework are the varied graph query languages and effective optimization techniques.
% This section introduces the existing query languages and optimization techniques, then presents the existing graph databases for optimizing \cgps.

\subsection{Graph Query Languages}
Numerous graph query languages \cite{gremlin,neo4j,gql,gcore,pgql,gsql} have been developed for querying property graphs~\cite{property_graph}. For instance, Gremlin~\cite{gremlin} represents graph traversal as a sequence of steps, {each offering functions such as navigating through graph vertices and edges}. % or retrieving their properties.
{Cypher \cite{cypher} crafted for Neo4j \cite{neo4j},
%stands out as one of the most prevalent graph query languages in academic and industry settings,
allows users to describe graph patterns with an ASCII-art syntax while integrating relational operators similar to those
%found
in SQL}. %G-Core \cite{gcore}, bearing resemblance to Cypher in syntax, strides towards composability where both the input and output of queries are graphs.
%Similarly, PGQL \cite{pgql} adheres to the SQL paradigm, extending it with capabilities for graph pattern matching, path finding, and reachability queries. GSQL \cite{gsql}, designed for TigerGraph \cite{tigergraph}, supports user-defined functions and procedures, enabling complex graph workloads that go beyond pattern matching ({e.g., PageRank and SSSP}).
%, including algorithms like PageRank and SSSP.
The recently published GQL \cite{gql} is a new standard developed by ISO committee for querying property graphs, with a consensus effort that integrates previous advancements.
% {More typical graph query languages like G-Core \cite{gcore}, PGQL \cite{pgql} and GSQL \cite{gsql} are not discussed here due to space constraints.}
{While these languages focus on querying property graphs, SPARQL \cite{sparql} is a W3C recommendation for querying Resource Description Framework (RDF) graphs \cite{rdf}.}
Despite the proliferation of graph query languages, there is a notable gap: the lack of a dedicated optimizer to enhance query efficiency in these languages.
%Despite the proliferation of these graph query languages, there remains a notable gap: the lack of a dedicated graph optimizer designed to enhance the efficiency of queries written in these languages.

\subsection{Graph Query Optimizations}
Graph pattern matching is fundamental in graph query processing.
%with subgraph isomorphism \cite{ullmann1976algorithm} being widely recognized as the most prevalent semantics.}
%Graph pattern matching is a fundamental problem in graph query processing and has been studied in numerous works. Subgraph isomorphism \cite{ullmann1976algorithm} is widely recognized as the most prevalent semantics for graph pattern matching.
In a sequential context, Ullmann's introduction of the first backtracking algorithm \cite{ullmann1976algorithm} marked a significant advancement in graph pattern matching. This has led to various optimizations, including tree indexing \cite{shang2008quicksi}, symmetry breaking \cite{han13turbo}, and compression techniques \cite{bi2016efficient}.
Due to the challenges in parallelizing backtracking algorithms, join-based algorithms have been developed for distributed environments. These algorithms often hinge on cost estimation to optimize join order for matching patterns efficiently. For instance, binary-join algorithms \cite{lai2015scalable, lai2019distributed} estimate costs using random graph models. In contrast, the work in \cite{ammar2018distributed} adopts the \wcoj~algorithm \cite{ngo2018worst}, ensuring a cost that remains within a worst-case upper bound.
{Recognizing the limitations of both binary join and \wcoj~methods in consistently delivering optimal performance~\cite{lai2019distributed}, researchers have turned to hybrid approaches \cite{aberger2017emptyheaded, mhedhbi2019optimizing, yang2021huge} that adaptively select between binary and \wcoj~techniques based on the estimated costs.}
%Recognizing the limitations of both binary join and {wcoj} methods in consistently delivering optimal performance~\cite{lai2019distributed}, researchers have turned to hybrid approaches \cite{aberger2017emptyheaded, mhedhbi2019optimizing, yang2021huge}. These approaches aim to adaptively select between binary and {wcoj} techniques based on which method incurs lower costs in a given situation.
For more accurate cost estimation, \cite{mhedhbi2019optimizing} and \cite{glogs} suggested to leverage high-order statistics.
%To further enhance cost estimation, the authors in \cite{mhedhbi2019optimizing} and \cite{glogs} have suggested leveraging high-order statistics to compute more effective execution plans for graph pattern matching.
While these studies have significantly improved graph pattern matching, they are not designed to address graph queries in practice that often encompasses relational operations. %their applicability is predominantly confined to that specific context. %They are not designed to address the broader spectrum of graph queries, which often encompass relational operations such as aggregation and projection.
{There is extensive work on RDF graph query optimization \cite{DBLP:conf/edbt/TsialiamanisSFCB12,DBLP:conf/www/StockerSBKR08, DBLP:conf/edbt/RabbaniLH21, 7930124}, which is not detailed here as this paper focuses on property graphs}.

% {
% For optimizing queries on RDF graphs, \cite{DBLP:conf/edbt/TsialiamanisSFCB12} outlines several heuristic rules designed to enhance SPARQL query optimizers,
% \cite{DBLP:conf/www/StockerSBKR08, DBLP:conf/edbt/RabbaniLH21, 7930124} propose cost-based optimization techniques based on cardinality estimation and join reordering,
% and \cite{DBLP:conf/edbt/YakovetsGG15} delves into optimizing regular path queries (RPQs) by utilizing fixpoint evaluation for property paths.
% This work focuses on query optimizations on property graph databases.
% }

\begin{table}
    \small
    \centering
    \caption{Limitations of Existing Graph Databases. ``Lang.'', ``Opt.'', ``H.~Stats'', and ``T.~Infer'' represent ``Language'', ``Optimization'', ``High-order Statistics'', and ``Type Inference''.}
    \label{tab:limitations}
    \begin{tabular}{ p{0.15\linewidth} | p{0.13\linewidth} | p{0.12\linewidth} | p{0.13\linewidth} | p{0.11\linewidth} | p{0.11\linewidth}}
    \hline
      \bf{Database} & \bf{Lang.} & \bf{Opt.} & \bf{WcoJoin} & \bf{H.~Stats.}& \bf{T. Infer} \\
      \hline
      Neo4j & Cypher & RBO/CBO & $\times$ &  $\times$  & $\times$ \\
      \hline
      GraphScope & Gremlin & RBO & $\checkmark$ & $\times$  & $\times$ \\
      \hline
      GLogS & Gremlin & CBO & $\checkmark$ & $\checkmark$ & $\times$ \\
      \hline
      \gopt & Multiple & RBO/CBO & $\checkmark$ & $\checkmark$ & $\checkmark$ \\
      \hline
    \end{tabular}
\end{table}

\subsection{Databases for Optimizing \cgps}
Graph databases~\cite{tigergraph, neo4j, glogs,janusgraph,neptune, graphflow} enable users to perform complex graph queries using declarative query languages.
However, they encounter limitations when it comes to fully supporting \cgps.
For example, Neo4j, a leading graph database, is limited by its tight coupling with Cypher~\cite{cypher} for \cgps~support.
GraphScope~\cite{graphscope}, a distributed graph computing system coupled with the Gremlin, lacks CBO optimization techniques.
%While \gs~supports worst-case optimal join, it only utilizes a rule-based optimizer for optimizing
%the query, which is not sufficient for complex patterns which requires cost-base optimization.
\glogs~\cite{glogs} is designed specifically to optimize \bgps,
ensuring worst-case optimality as well as leveraging high-order statistics (i.e., the small pattern frequencies), but it cannot optimize \cgps~due to its inability to handle relational operations or arbitrary types.
More limitations are summarized in \reftab{limitations}.
{Note that Neo4j does not perform type inference because of its schema-loose design for greater flexibility. }
In summary, current graph databases face challenges in optimizing \cgps~ effectively.

% Neo4j faces challenges with pattern matching execution, as it cannot ensure worst-case optimality \cite{ngo2018worst, glogs}, potentially resulting in suboptimal performance. Additionally, its schema-loose design limits type inference capabilities, compelling Neo4j to interpret pattern graph vertices or edges with \unionall~as matching all possible vertices or edges within the data graph, which can be inefficient.
% On the other hand, GraphScope~\cite{graphscope} is a distributed graph computing system integrated with the Gremlin query language. Although GraphScope supports worst-case optimal joins, it relies solely on a rule-based optimizer for query optimization. This approach is inadequate for complex patterns that necessitate a cost-based optimization strategy for efficient execution.
% \glogs~\cite{glogs} presents a cost-based optimizer tailored to enhance the performance of \bgps~matching. To achieve more precise cardinality estimations within its optimization process, it pre-computes small pattern frequencies, termed as high-order statistics. However, \glogs~falls short in offering native support for relational operators in \cgps and lacks optimization concerning \utypes. In summary, current graph databases face challenges in optimizing \cgps~ effectively.

A \emph{non-graph-native} approach involves converting \cgps~ into relational queries and leveraging relational optimizers, as seen in systems like IBM DB2 Graph\cite{DBLP:conf/sigmod/TianXZPTSKAP20}, SQLGraph~\cite{DBLP:conf/sigmod/SunFSKHX15}, and others~\cite{hassan18, jin2022, relgo}. The primary advantage of this method lies in its ability to build upon the well-studied relational optimization techniques.
However, this approach fails to utilize graph-specific optimizations \cite{mhedhbi2019optimizing, yang2021huge, glogs} and struggles to handle \utypes~(i.e., a combination of multiple types, detailed in \refsec{preliminaries}) in \cgps, a construct inherently tied to graph semantics and absent in relational models, leading to inefficient query execution.
% For instance, to match a pattern with \utypes~in \reffig{workflow}(a), it might have to execute multiple queries, each targeting a pattern where every vertex and edge is assigned with a specific type, and then merge the results, which is inefficient.}
For example, matching the pattern with \utypes~in \reffig{workflow}(a) may require executing multiple queries—each with specific types for pattern vertices and edges—and then merging the results, leading to inefficiency.

\vspace*{-1ex}
\section{Preliminaries}
\label{sec:preliminaries}
%We present some background and elaborate the challenges in processing \patrel~in this section.
% {In this section, we present the definitions of notations used in this paper and introduce the limitations of exisiting systems.}

\begin{table}[t]
  \small
  \centering
  \caption{Frequently used notations.}
  \label{tab:notations}
  \begin{tabular}{ p{0.2\linewidth} | p{0.7\linewidth}  }
  \hline
    \bf{Notation} & \bf{Definiton} \\
    \hline
    $G(V_G, E_G)$ & A data graph with $V_G$ and $E_G$ \\
    \hline
    $P(V_P, E_P)$ & A pattern graph with $V_P$ and $E_P$ \\
    \hline
    $\footnotesize{\nbr{G}}(v)$, $\footnotesize{\nbre{G}}(v)$ & Out neighbors and out edges of $v$ in graph $G$ \\
    \hline
    $\elemtype_G(v)$, $\elemtype_G(e)$ & The type of vertex $v$ and edge $e$ in graph $G$ \\
    \hline
    $\type_P(v)$, $\type_P(e)$ & The type constraint of vertex $v$ and edge $e$ in pattern $P$, can be \btype, \utype~or \unionall\\
    \hline
    $\freq_{P,G}$ & The number of mappings of pattern $P$ in graph $G$ \\
    \hline
  \end{tabular}
\end{table}

% \subsection{The Definitions}
% \label{sec:patrel}
{Data graph $G = (V_G, E_G)$ in this paper adheres to the definition of the property graph model \cite{angles17}.
$V_G$ and $E_G$ are the sets of vertices and edges, where $|V_G|$ and $|E_G|$ represent the number of vertices and edges. }
% We adopt a property graph model~\cite{angles17}  to represent the data graph.
% The graph is defined as $G = (V_G, E_G)$,
% where $V_G$ and $E_G$ are the sets of vertices and edges.
% %where $V_G$ and $E_G$ are the sets of vertices and edges in $G$ respectively.
% We use $|V_G|$ and $|E_G|$ to denote the number of vertices and edges respectively.
% Graph size $|G|$, is defined as the number of vertices in the graph.
{Given $u, v \in V_G$, $(u, v) \in E_G$ is an edge directed from $u$ to $v$, and all the out neighbors and out edges of $u$ are denoted as $\nbr{G}(v)$ and $\nbre{G}(v)$, respectively.}
% Each edge $(u,v)\in E_G$ is directed from $u$ to $v$.
% %Each vertex $v\in V_G$ represents an entity,
% %and each edge $(u,v)\in E_G$ represents a directed relationship from $u$ to $v$.
% For each vertex $v$, $\overrightarrow{\nbr{G}}(v)$ (resp. $\overleftarrow{\nbr{G}}(v)$) and $\overrightarrow{\nbre{G}}(v)$ (resp. $\overleftarrow{\nbre{G}}(v)$) represents the out neighbors and out edges (resp. in neighbors and in edges) of $v$ in $G$.
%For each vertex $v$, we use $\overrightarrow{\nbr{G}}(v)$ (resp. $\overleftarrow{\nbr{G}}(v)$) and $\overrightarrow{\nbre{G}}(v)$ (resp. $\overleftarrow{\nbre{G}}(v)$) to denote the out neighbors and out edges (resp. in neighbors and in edges) of $v$ in $G$.
% Each vertex or edge in $G$ is associated with a type to represent its class,
{Each vertex or edge in $G$ is associated with a type,}
denoted as $\elemtype_G(v)$ and $\elemtype_G(e)$ respectively.
Both vertices and edges can carry properties, which are key-value pairs.
Note that if no ambiguity arises, we omit $G$ from the subscript in the notations, and so as the follows.
%We use $\prop(v)$ and $\prop(e)$ to denote the set of properties of vertex $v$ and edge $e$ respectively.
{Besides, although our framework supports multi-type vertices as in \cite{cypher}, this paper focuses on vertices with single type for simplicity}.
{Considering two graphs \(G_1\) and \(G_2\), we assert that \(G_2\) is a subgraph of \(G_1\), symbolized as \(G_2 \subseteq G_1\), if and only if \(V_{G_2} \subseteq V_{G_1}\), and \(E_{G_2} \subseteq E_{G_1}\).
%Furthermore, \(G_2\) qualifies as an induced subgraph of \(G_1\) under the condition that \(G_2\) is already a subgraph of \(G_1\), and for every pair of vertices in \(G_2\), any edge \(e\) that exists between them in \(G_1\) must also present in \(G_2\).
}
% {For simplicity, we do not discuss the case where a vertex can be associated with multiple types, as in~\cite{cypher}. While our framework supports this functionality, it lies outside the scope of this paper.}

\eat{
{
The graph schema defines the types of vertices and edges in the data graph.
However, it can be different in schema-strict and schema-loose graph databases.
In a schema-strict graph database \todo{citations, e.g., KuzuDB, GRainDB, TigerGraph}, which means you have to define the schema first, and all data must conform to the schema.
Thus in this case, the schema is predefined and remains static.
In contrast, a schema-loose graph database indicates a dynamic schema \todo{citations, e.g., Neo4j, JanusGraph. JanusGraph supports both schema-strict and schema-loose}, where the schema can be updated during the data insertion or updating process.
In this cases, the schema can be predefined but the data can be not strictly conform to the schema (i.e., the schema can be updated), or the schema can be empty at the beginning and gradually uncovered during the data insertion or updating process.
To support both schema-strict and schema-loose graph databases, we introduce an optional vertex in the schema to represent the unknown vertex types in the data graph when it is schema-loose, denoted as \xtype.
\xtype has three states: no updates, partial updates, and unknown updates.
Assume we have an initial schema $S = (V_S, E_S)$, where $V_S$ and $E_S$ are the sets of vertex and edge types respectively.
When dealing with schema-loose graph databases,
the updated schema is defined as $S = (V_S\cup \{\xtype\}, E_S \cup \{(v, \xtype) | \forall v\in V_S\cup\{\xtype\}\})$.
During the data insertion or updating process, we batch the vertices and edges, and gradually uncovering the specific types denoted by \xtype.
}
}

A pattern $P = (V_P, E_P)$ is a small \textit{connected} graph.
Note that if $P$ is not connected, matching the pattern is equivalent to taking the Cartesian product of
the matches for its connected components, which is naturally the problem of \cgps~ in the following.
Each vertex and edge in the pattern graph is associated with a set of types as a type constraint, denoted as $\type_P(v)$ and $\type_P(e)$, respectively.
%Following \cite{Angles_2023},
{Three categories of type constraints are supported: }
% We support three categories of type constraints:
(1) \btype~contains a single type that matches a particular vertex (or edge) {type} in the data graph.
(2) \utype~encompasses multiple types, allowing for a match with data vertices (or edges) of any type in this set.
{For instance, \utype \code{\{Post,Comment\}} signifies that the matched vertices can be either of type \code{Post} or \code{Comment} in data graph}.
% {For instance, the type constraint of \code{\{Post,Comment\}} signifies that the matched vertices can be either of type \code{Post} or \code{Comment} in data graph}.
(3) \unionall, considered as a special \utype, indicates that any vertex or edge in data graph satisfies the constraint. %For example, {all vertices and edges except $v_3$ in \reffig{workflow}(a)} are initialized with type constraints \unionall.
Predicates can be specified to vertices and edges in the pattern graph as well, while we mostly focus on the type constraints in this paper.

Finding matches of $P$ in $G$ involves identifying all subgraphs $G'$ in $G$ {where} $P$ can be mapped to $G'$ via a homomorphism preserving edge relations and type constraints.
{The mapping function $h: V_P \rightarrow V_{G'}$ ensures that $\forall e=(u, v) \in E_P$, there is a corresponding edge $(h(u), h(v)) \in E_{G'}$}.
Additionally, the types of vertices and edges in $G'$ must align with the type constraints specified in $P$:
{(1) $\forall v \in V_P, \elemtype_{G'}(h(v)) \in \type_P(v)$, and (2) $\forall e=(u, v) \in E_P, \elemtype_{G'}((h(u), h(v))) \in \type_P(e)$.}
The number of mappings of $P$ in $G$ is called \textit{pattern frequency}, {denoted as $\freq_{P,G}$, or simply $\freq_P$ when $G$ is clear.}
\begin{remark}
\label{rem:semantics}
The homomorphism semantics, which allows duplicate vertices or edges in matching results, is employed in our framework due to its transformability. 
Specifically, a plan using homomorphism semantics can be converted to a plan using other commonly adopted semantics, such as the no-repeated-edge semantics (used by Cypher) that excludes duplicate edges, by adding an all-distinct filter at the end of the \bgp~match. 
This makes our framework flexible and adaptable to different query language semantics.
% {The homomorphism semantics permits duplicate vertices or edges in matching results, and is used in our framework due to its transformability.
% That is, given another commonly used semantics \cite{angles17} such as no-repeated-edge semantics used by Cypher, which excludes results with duplicate edges, a plan in homomorphism semantics can be converted to a plan in that semantic easily by adding an all-distinction filter at the end of the \bgp~ match.
% This feature enhances our framework's generality and flexibility, enabling integration with different semantics in query languages.}
\end{remark}

{A complex graph pattern, termed as \cgp, extends \bgps~with further relational operations}.
{Optimizing \cgps~is challenging due to their hybrid semantics.
A straightforward way is to first identify matches of $P$ in $G$, and then subject the matched subgraphs to the remaining relational operators in \cgps~ for further analysis, such as projecting the properties of matched vertices and edges and selecting the matched subgraphs that satisfy certain conditions.}

\vspace*{-1mm}
\section{System Overview}
\label{sec:arch}

\begin{figure}[t]
    \centering
    \includegraphics[width=0.65\linewidth]{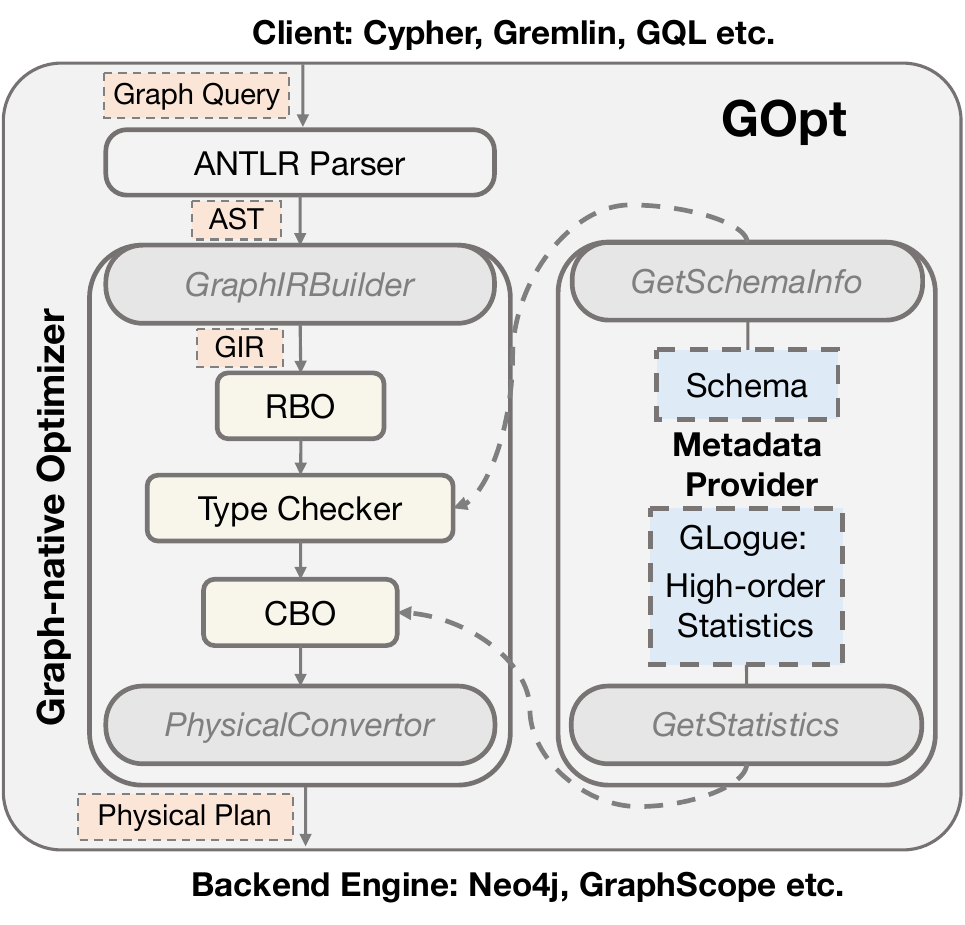}
    \vspace*{-0.5em}
    \caption{System Architecture Overview}
    \label{fig:arch}
\end{figure}

\begin{figure*}[t]
    \centering
    \includegraphics[width=0.9\linewidth]{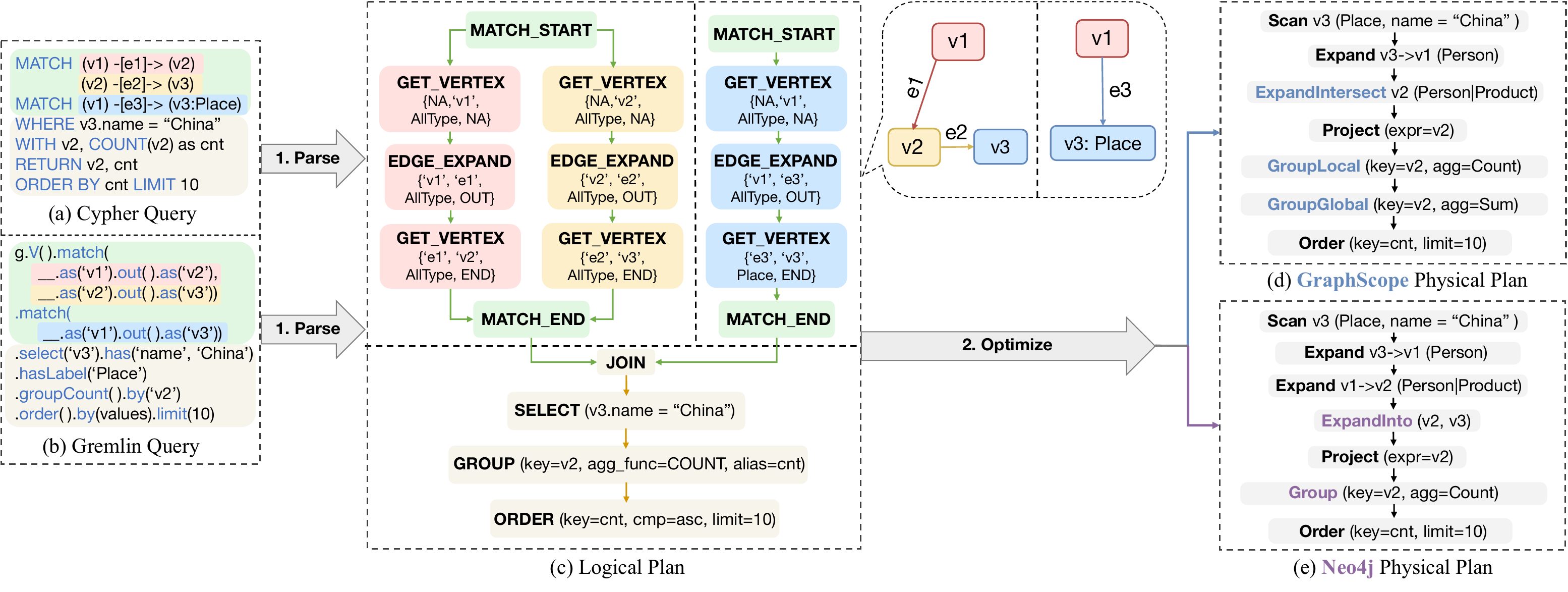}
    \caption{An example of query processing workflow. For simplicity, we draw the pattern graph for \matchpattern~(examples are given) in the following. In this paper, logical operators are in ALL\_UPPERCASE format, and physical operators use CamelCase.}
    \label{fig:workflow}
\end{figure*}

{To address the limitations in existing systems, we propose a graph-native query optimization framework, \gopt, {specifically for \patrels}. This framework leverages state-of-the-art graph query optimization techniques (with refinement and extension for \cgps), offers a unified approach for supporting multiple query languages, and facilitates seamless integration with various backend engines.}
The system architecture is shown in \reffig{arch}, {with} {three} principal components of query parser,
optimizer, and metadata provider.
% Within this framework, the \gopt, which serves as the frontend component, is the primary focus in this paper.
%\subsection{System Architecture}

% \stitle{Client.} The client component is the entry point for users,
% providing an interface for submitting queries to the system and for retrieving the results.
% It is important to note that the system accommodates various clients for different graph query languages, such as Gremlin and Cypher.
% Moreover, the system is designed to maintain compatibility with GQL, the emerging standard for graph query languages,
% ensuring its relevance and adaptability to future developments in the field.

% \stitle{Frontend.} The frontend component acts as a vital bridge between the client and the backend.
% Specifically, \gopt~which we mainly focus on in this paper,
% serves as the frontend module within the whole system.
% \gopt encompasses three main sub-modules:

\stitle{Query Parser}.
{\gopt~ employs the ANTLR parser tool~\cite{antlr} to generate an AST for queries in various graph query languages, such as Gremlin and Cypher. Then it utilizes a \graphirbuilder~to convert the AST} into a \textit{language-independent} plan based on a unified graph intermediate representation (\ir), detailed in \refsec{ir}.
This \ir~decouples \gopt~from specific query languages and allows extensive reuse of techniques developed within \gopt.
%Query parser plays a pivotal role in transforming queries into a \textit{language-independent} query plan based on the unified intermediate representation (IR) detailed in \refsec{parser}.
% It supports various clients for different graph query languages, such as Gremlin and Cypher.
%and is designed to maintain the compatibility with ISO/GQL, the emerging standard for graph query languages.

\eat{
\stitle{Type Checker}.
Type checker is responsible for inferring and validating the type constraints in the query pattern against the graph schema.
This is particularly significant when the pattern lacks explicit type constraints.
The basic idea is to iteratively infer type constraints for the vertices and edges in pattern,
based on the user-specified type constraints and schema-indicated candidate types.
During the process, if no more valid types can be assigned, the type checker will return an INVALID flag as the pattern fails in type checking.
The type checker is critical in enabling the system to handle \arbitrary~in query patterns, ensuring implicit type constraints are upheld. This significantly improves cost estimation accuracy during the optimization phase and boosts execution efficiency.
}

\stitle{Optimizer}.
The graph-native optimizer is the core module in \gopt, employing a combination of Rule-based Optimization (RBO), type checker, and Cost-based Optimization (CBO).
% {The default optimization process follows this sequence: RBO is applied first, followed by type inference, and finally CBO. This order is designed to balance optimization effectiveness with computational cost, recognizing that both RBO and type inference can influence the outcomes of CBO.}
%{When optimizing with GOpt, these components are applied sequentially to enhance the CGPs.}
The RBO comprises a comprehensive set of heuristic rules designed to optimize the interactions between \bgps~and relational operations in \cgps.
Apart from the rules originally included in RBO, new rules can be integrated to further enhance its optimization capabilities.
However, since RBO does not account for data characteristics, it may fail to fully {optimize \cgps}, particularly in the presence of complex \bgps. Therefore, we need to explore additional data-driven optimization techniques for handling \cgps.

{The first such technique is a type checker, specifically designed to infer implicit type constraints from the underlying data for \bgps~with arbitrary types.
%The type checker is applied after RBO because some RBO rules, such as \joinelimrule that merges pattern graphs, provide more connectivity information for type inference, thereby enhancing its accuracy.
This technique is crucial for query optimization, enhancing execution efficiency and improving cardinality estimation accuracy in the subsequent CBO phase}.
% This technique plays a critical role in query optimization, as it not only enhances execution efficiency but also improves the accuracy of cardinality estimation during the subsequent CBO phase}.

Building on this, we propose a unique CBO approach for further optimization.
 Our CBO introduces several key innovations for \bgps. First, we leverage high-order statistics from \glogs~\cite{glogs} for more accurate estimation of \bgp~ frequencies. Unlike \glogs, however, our approach handles arbitrary type constraints.
Second, to accommodate diverse backend engines, we allow backends to register their specific physical operators and cost models through a \physicalbuilder.
Finally, we introduce a top-down search algorithm with a branch-and-bound strategy to determine the optimal query plan, aiming to minimize the estimated cost. Together, these techniques ensure robust and adaptable query optimization across various systems.
We build \gopt~based on Apache Calcite~\cite{Begoli_2018},
a prominent open-source optimization framework for relational databases, to leverage Calcite's relational optimization techniques.

\stitle{Metadata Provider}.
{The metadata provider has two key components: the first is the graph schema, defining the types of vertices and edges in the data graph for the type checker.
For simplicity, we initially assume a schema-strict context with an explicit schema,
and more flexible contexts are discussed later in \refrem{schema-loose}}.
The second component, named \glogue following~\cite{glogs}, serves as a data statistics provider.
\eat{
The metadata provider is responsible for supplying the optimizer with the necessary metadata for cost estimation.
An important component is \glogue~\cite{glogs}, which is a high-order statistics provider.
}
It precomputed the frequency of small graph patterns (a.k.a., motifs) {with up to $k$ vertices ($k \geq 3$)}
%(where $k$ is at least 3)
in the data graph, going beyond the low-order statistics which only count the vertices and edges.
These advanced high-order statistics afford the optimizer with more precise cardinality estimation for query patterns, thereby enhancing the effectiveness of the optimization.
% \eat{
% At the beginning of the system initialized,
% \glogue~is built from scratch to encompass all possible patterns up to a certain size, defined by $k$ vertices (where $k$ is at least 3),
% composed with types defined in the graph schema, whereas the patterns are limited to \bpatterns~only.
% \glogue~then precomputes the frequencies of each pattern in the data graph,
% maintaining them as high-order statistics.
% Although \glogue~specifically addresses \bpatterns~and cannot directly provide high-order statistics for \upatterns,
% it can still be leveraged as a groundwork for approximating the high-order statistics of arbitrary patterns during optimization.
% The details of \glogue~are beyond the scope of this paper and are therefore not discussed here.
% }

\stitle{Overall Workflow.} {The query processing workflow of \gopt~is illustrated in \reffig{workflow}. After parsing the queries, the optimization process proceeds in the following order: RBO, type inference, and finally CBO. This sequence is designed to balance optimization effectiveness with computational cost, as both RBO and type inference can influence CBO outcomes.
Details on query parsing, heuristic rules, type inference, and cost-based optimizations are provided in \refsec{trans}, \refsec{rbo}, \refsec{typeinfer}, and \refsec{cbo}, respectively. This paper focuses specifically on optimizing graph \bgps. For instance, in CBO, our primary focus is on the costs associated with these \bgps. The handling of relational operators and their costs is further detailed in \refrem{impl_calcite}.  %For example, in components like CBO, our concern is limited to the costs associated with these patterns. Once the graph patterns are optimized, they can be encapsulated within a specialized relational SCAN operator. This allows the entire plan to be converted into a purely relational one, and if needed, using existing relational techniques for further optimization, as will be discussed in \refsec{impl}.
    %, where the cost is determined by the pattern frequencies. This allows the entire plan to be converted into a purely relational one, enabling further optimization, if needed, using existing relational techniques, as will be discussed in \refsec{impl}.
}

% \stitle{Backend.} The backend component is responsible for executing the optimized query plan and returning the query results to the user.
% This component can function on a standalone machine or be distributed across a cluster of machines,
% depending on the data scale and specific requirements of the system.
% As the backend is not the focus of this paper,
% we integrate an existing dataflow engine \gaia~as the backend for our system.
% To make the physical plan executable in \gaia,
% we have incorporated a dataflow plugin library.
% This library includes code generation capabilities for the operators in the physical plan,
% along with a job assembler designed to assemble the dataflow job.
% In addition, we utilize an in-memory graph store that supports partitioned graph, to enable efficient access to graph data.

\section{Query Transformation}
\label{sec:trans}

% In this section, we first introduce the intermediate representation (IR) abstraction for \patrels.
% Then we describe how \gopt~converts queries in various graph query languages into a unified logical plan based on IR.
% After that a phase of type inference and validation is applied to confirm that the plan's type constraints comply with the graph schema.

To allow different query languages to plug into \gopt, we present the {unified} graph intermediate representation (\ir) for \patrels~and describe interfaces to convert queries into this format.
\eat{Then we introduce the method of type inference and validation to ensure that the plan's type constraints comply with the graph schema.}

\subsection{\ir Abstraction}
\label{sec:ir}
{The \ir~abstraction~\cite{DBLP:conf/sigmod/HeHLLLLLLLMSSW024} provides a foundation for representing \cgps~with both \bgps~and relational query semantics}.
It defines a data model $\mathcal{D}$ that presents a schema-like structure for intermediate results in which each data field has a name and a designated datatype, including graph-specific datatypes (e.g., \textit{Vertex}, \textit{Edge}, \textit{Path}) and general datatypes (e.g., \textit{Primitives} and \textit{Collections}).
It also defines a set of operators $\Omega$ that operate on data tuples from $\mathcal{D}$ and produce new data tuples as results, consisting of graph operators and relational operators.
{The graph operators are designed to retrieve graph data, including: (1) \expandedge~to expand adjacent edges from vertices; (2) \getvertex~to retrieve endpoints from edges; (3) \expandpath~to support path expansion with a specified hop number $l$, {corresponding to $l$ consecutive edges in the pattern,} and users can also define path constraints such as \emph{Arbitrary} (no constraints), \emph{Simple} (\emph{No-repeated-node}), or \emph{Trail} (\emph{No-repeated-edge})~\cite{angles17}; %The \emph{Simple} and \emph{No-repeated-edge} constraints are particularly useful for preventing unbounded path exploration; 
and (4) \matchpattern~as a composite operator that encapsulates the above graph operators between \matchstart~and \matchend~to represent a \bgp, as illustrated in \reffig{workflow}(c). 
The relational operators consist of the commonly used operators in RDBMS and can be applied on graph data, e.g., \project~for projecting vertex properties, \select~for selecting edges based on conditions, \joinopr~for joining sub-paths, etc. More details of the \ir~ can be found in~\cite{artifact}.
}

\eat{
The graph operators are designed to retrieve graph data, including \expandedge~to expand adjacent edges from vertices,
\getvertex~to retrieve endpoints from edges,
{\expandpath~to expand paths with a specified hop number $l$, where users are also allowed to define path options such as to match a simple path with no repeated vertices or edges,}
and {\matchpattern~as a composed operator, enclosed by \matchstart~and \matchend, denoting \bgps~in \ir~format}.

The relational operators include the commonly used operators in RDBMS and can be applied on graph-specific data as well, e.g., \project~for projecting vertex properties, \select~for selecting edges with specific conditions, \joinopr~for joining two sub-paths into a longer one, \union~for merging the results of two sub matching clauses, etc.
{More details of the IR definition can be found in \cite{artifact}.} %{An example of an IR ?}
}

\begin{remark}
  {The definition of \ir~is inspired by existing work on graph relational algebra~\cite{Szrnyas2018ReducingPG}
  %but we extend it by introducing \matchpattern~to enable the processing of complex \bgps. 
  {and is enhanced with \matchpattern~to handle complex \bgps}.
  Moreover, \ir's development~follows an engineer-oriented approach, prioritizing support for commonly used functionalities over theoretical completeness. It {continuously} evolves to meet new requirements from real applications.}
\end{remark}

With the \ir abstraction, a \patrel~is represented as a unified directed acyclic graph (DAG),
where the nodes represent operators in $\Omega$, and the edges define the data flow between operators.
The DAG has two representations, the \textit{logical plan} for high-level query semantics (shown in \reffig{workflow}(c)) and the \textit{physical plan} for low-level operator implementations (shown in \reffig{workflow}(d) and \ref{fig:workflow}(e)).
To construct the logical plan, \gopt~offers a \graphirbuilder, as exemplified in \refsec{parser}.
For the physical plan, where operators are tied to the backend engine, \gopt~uses a \physicalbuilder~to allow backends to register their own physical operators, ensuring efficient plan execution. Physical plans on two different backends are exemplified in \reffig{workflow}(d) and \reffig{workflow}(e), and further details are provided in \refsec{cbo}.
%given that the physical operators are closely related to the underlying backend engine, \gopt~employs a \physicalbuilder~to allow the backends to register their specific physical operators to ensure that the plan is executed correctly and efficiently. Examples of two physical plans are shown in \reffig{workflow}(d) and \reffig{workflow}(e), and further details {are provided} in \refsec{cbo}.

\subsection{Query Parser}
\label{sec:parser}
\gopt~employs the ANTLR parser tool~\cite{antlr} to interpret queries in different query languages into an Abstract Syntax Tree (AST).
Currently, \gopt~supports two of the most popular graph query languages, Cypher\cite{cypher} and Gremlin\cite{gremlin}, as exemplified in \reffig{workflow}(a) and \reffig{workflow}(b),
and preserves compatibility with the recent ISO standard GQL~\cite{gql}.
Then, the AST is transformed into the \ir~ as a logical plan, as shown in \reffig{workflow}(c),
through a \graphirbuilder.
A code snippet for building the logical plan is as follows:
\begin{lstlisting}
GraphIrBuilder irBuilder = new GraphIrBuilder();
pattern1 = irBuilder.patternStart()
  .getV(Alias("v1"),AllType())
  .expandE(Tag("v1"),Alias("e1"),AllType(),Dir.OUT)
  .getV(Tag("e1"),Alias("v2"),AllType(),Vertex.END)
  .expandE(Tag("v2"),Alias("e2"),AllType(),Dir.OUT)
  .getV(Tag("e2"),Alias("v3"),AllType(),Vertex.END)
  .patternEnd();
pattern2 = irBuilder.patternStart()
  .getV(Alias("v1"),AllType())
  .expandE(Tag("v1"),Alias("e3"),AllType(),Dir.OUT)
  .getV(Tag("e3"),Alias("v3"),BasicType("Place"),Vertex.END)
  .patternEnd();
query = irBuilder.join(pattern1,pattern2,
    Keys([Tag("v1"), Tag("v3")]), JoinType.INNER)
  .select(Expr("v3.name='China'"))
  .group(Keys(Tag("v2")),AggFunc.COUNT, Alias("cnt"))
  .order(Keys(Tag("cnt")), Order.ASC, Limit(10));
\end{lstlisting}
{%which means, we retrieve all vertices with \unionall~as $v_1$, expand all out edges from $v_1$ as $e_1$, and retrieve the target vertex $v_2$ from $e_1$.
Note that \code{Alias()} defines an alias for results, accessible via \code{Tag()} in later operations. With the \graphirbuilder, the logical plan is constructed in a language-independent manner, facilitating the subsequent optimization process. Henceforth, when optimizing CGPs, we are referring to optimizing their GIR form.
}
\eat{
{In the logical plan, we further construct a \pattern~in graph structure for \matchpattern, where \getvertex~corresponds to vertices, \expandedge~corresponds to edges, and \expandpath~with hop number $l$ corresponds to $l$ consecutive edges in the pattern, to facilitate the subsequent process.
For example, the \matchpattern~in \reffig{workflow}(c) is further built into a \pattern~in \reffig{typeinfer}(a), based on which the type inference and optimizations are performed.
Specifically, when transforming the \expandpath, we preserve the path's start, end, and options. This preservation ensures that, after the optimization process, it can be transformed back into an \expandpath, or possibly join of \expandpath~operators depending on the optimization results, within the physical plan to maintain the original path matching semantics.}
}

\eat{
\begin{example}
%\reffig{workflow}(a-c) illustrates the process of parsing a query into a logical plan.
\reffig{workflow}(a-b) illustrate the original queries written in Cypher and Gremlin respectively.
In the Cypher query, we parse \code{MATCH (v1)-[e1]->(v2)} into $\getvertex(NA, v_1,\unionall, NA)$, $\expandedge(v_1, e_1, \unionall, OUT)$, and then followed by another $\getvertex(e_1, v_2, \unionall, TGT)$.
Similar process are applied to subsequent clauses in the \code{MATCH}, resulting in the \matchpattern~\\shown in the upper portion of \reffig{workflow}(c).
The \code{match} part in Gremlin query can be parsed into a \matchpattern~in a similar way.
For the relational part, e.g., the \code{WHERE} clause in Cypher, is parsed into a \select~operator with the filter condition of \code{v3.name="China"}.
Finally, the logical plan is obtained as depicted in \reffig{workflow}(c).
\end{example}
}

\section{Optimization}
\label{sec:opt}
{In this section, we present our optimization techniques, including RBO with carefully designed heuristic rules for \cgps~(\refsec{rbo}), a type inference method for arbitrary \bgps~(\refsec{typeinfer}), and CBO techniques (\refsec{cbo})}.
% In this section, we present our optimization techniques, including rule-based optimization (RBO) with four carefully designed heuristic rules for \cgps~(\refsec{rbo}), a type inference method for arbitrary \bgps~(\refsec{typeinfer}), and cost-based optimization (CBO) techniques (\refsec{cbo}).

\eat{
\subsection{Optimization Framework}
\label{sec:optframework}
\gopt~is designed to optimize \patrels~in a unified way.
The overall optimization process is shown in \reffig{opt_workflow}.
Given a \patrel~$Q$, \gopt~first employs Rule-Based Optimizations (RBO), as shown in \reffig{opt_workflow}(b).
A series of rules have been devised, encompassing relational rules, graph rules, and relational-graph interplay rules, to optimize $Q$ in a heuristic way.
% For example, \filterrule~is a graph-relational rule to push the filter condition in \select~into the corresponding vertices and/or edges in the pattern,
% in order to reduce the intermediate results produced in the patten matching stage.
%
However, RBO relies solely on the predefined rules and does not take the statistical characteristics of the data into consideration.
To address this limitation, we further introduce Cost-Based Optimizations (CBO), illustrated in \reffig{opt_workflow}(c).
Specifically, the CBO tailored for optimizing \patrels~consists of the graph optimization phase and the relational optimization phase.
The first phase focuses on optimizing pattern matching to find an efficient search order.
Recall that in \cite{glogs}, the authors proposed \glogs~system aiming to optimize pattern matching based on \glogue which preserves high-order statistics for patterns.
However, \glogue~is limited to \bpatterns~and does not support \upatterns.
To overcome this limitation, we introduce a novel approach to estimating pattern cardinality by taking arbitrary type constraints into account, so as to support arbitrary patterns in \patrels.
Once the search order for pattern has been established, we proceed to the second phase, the relational optimization stage.
In this phase, our goal is to further refine the query execution plan by incorporating existing relational optimization techniques.
For example, as shown in \reffig{opt_workflow}(c), we optimize the \group~operator in distributed execution,
by first performing group operations locally, followed by a subsequent consolidation at the global level,
so as to minimize the communication costs during the grouping process.
}

\begin{figure}[t]
    \centering
    \includegraphics[width=0.95\columnwidth]{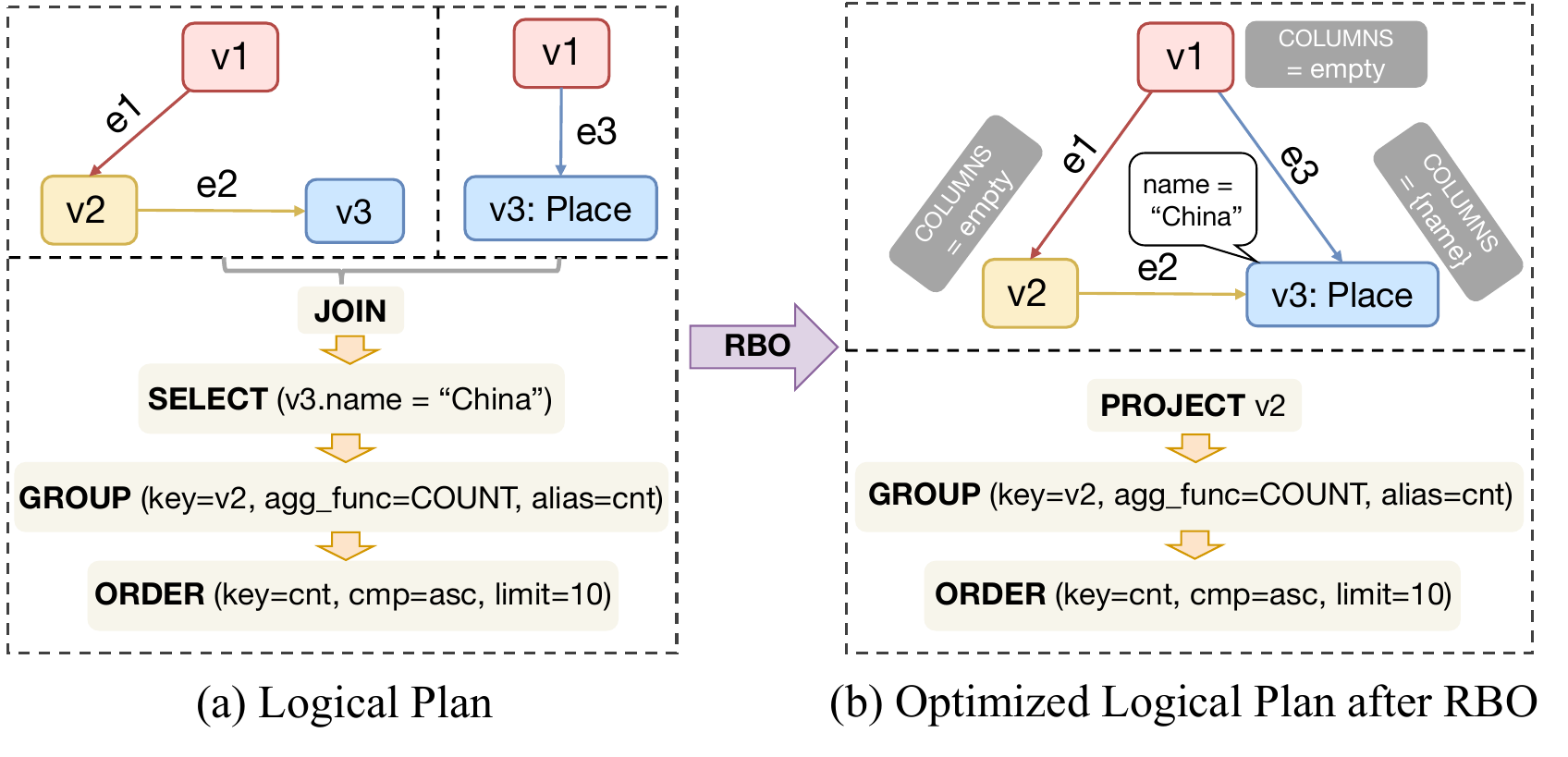}
    \caption{An Example of Rule-Based Optimizations}
    \label{fig:rbo}
\end{figure}

%By combining RBO and CBO, which together address both graph-specific and relational optimizations, \gopt~is able to improve the performance of \patrel processing.
\vspace{-2ex}
\subsection{Rule-based Optimization}
\label{sec:rbo}
Given a \patrel~$Q$, \gopt~first employs Rule-Based Optimizations (RBO) to optimize $Q$ in a heuristic way.
\eat{These pre-defined rules consider possible optimizations among graph operators, relational operators, and across both.}
{These pre-defined rules consider possible optimizations across {\bgps}~and relational operators. Note that the heuristic rules are designed to be extensible and pluggable.}
We introduce four representative rules in the following.

\stitle{\filterrule.}
{Consider a \bgp~ followed by a \select~ operator}.
The \filterrule~aims to push filters into {\bgps}~to reduce the number of intermediate results.
\eat{This is a graph-relational rule aiming to push filters into graph operators as early as possible.}
Typically, querying {\bgp} in graph databases can result in numerous matchings.
Thus, the users usually apply filters with \select~operator to narrow them down.
% when querying graph databases, the {\bgp}~ is used to identify specific pattern matchings. However, these matchings can be numerous, thus users usually apply filters with the \select~operator to narrow them down.
This rule optimizes the process by integrating filters with graph operators during the pattern matching phase.
\eat{For example, in \reffig{opt_workflow}(b), the filter condition \code{name="China"} is pushed into $v_3$ to reduce the number of intermediate results.
This rule is akin to the traditional \selectpush~in relational databases, which pushes filter through joins, but it is specifically tailored for \patrels, thanks to the unified IR abstraction.}
{For example, the \select~operator in \reffig{rbo}(a) is pushed down into the pattern, applying constraints during matching $v_3$ as shown in \reffig{rbo}(b). }

% {For example, in \reffig{rbo}(a), \gopt~first leverages the relational \selectpush~rule that has already been developed in Calcite, which pushes filters after the corresponding {\bgps}, and \filterrule~further pushes filters into {\bgps}, after the specific {\expandedge}~and \getvertex~operators in it, as shown in \reffig{rbo}(b).}

\stitle{\trimrule.}
{Given a \bgp, the \trimrule~is designed to eliminate unnecessary intermediate data during query execution, which is useful when the aliased intermediate results are no longer needed in the subsequent query processing.
For example, in \reffig{rbo}(a), since only $v_3$'s \code{name} is used in the query, we specify retrieving just this attribute during pattern matching using \code{COLUMNS}.
Similarly, since the query focuses solely on $v_2$'s count after pattern matching, \trimrule~uses a \project~ operator to trim the other vertices and edges. }

% {For a {\bgp~followed by} a \project~or other operators that work on certain matched vertices, edges, or their properties from {the \bgp}, the \trimrule~is designed to eliminate unnecessary intermediate data during query execution, which is useful when the aliased intermediate results are no longer needed in the subsequent query processing.}
\eat{This is a traditional relational rule that eliminates unnecessary intermediate data during query execution. It is useful when:
(1) the aliased intermediate results are no longer needed in the subsequent query processing, and
(2) the properties of vertices and edges are not utilized in the query.
}
% For example, in \reffig{rbo}(a), the aliased matched vertices of $v_1$ and $v_3$ can be discarded using a \project~after pattern matching to retain only $v_2$.
% For properties, since only $v_3.$\code{name} is used in the subsequent filter condition, we specify this required property with a \code{COLUMNS} field containing ``name'' for $v_3$.

\eat{
\stitle{\fusionrule}. This is one of the graph rules that aims to combine \expandedge~and \getvertex~into a single \expandvertex,
since in most cases, we are searching for neighboring vertices, not the edges.
However, whether or not we can combine the two operators depends on specific situations.
For instance, if we are working in a distributed system and need to retrieve certain properties of the vertices with \getvertex,
we might not be able to fuse the operators since the neighbors' properties may not be available locally.
This rule takes such conditions into account to optimize the query while guarantees consistent results.
}

\stitle{\joinelimrule.}
{For two {\bgps}~connected by a \joinopr~operator, the \joinelimrule~aims to merge them into a single one, when the {join keys (i.e., vertices and/or edges)} serving as the common vertices and/or edges in the resulted {\bgp}.
{This rule is effective under the homomorphism-based matching semantics (\refrem{semantics})}.
However, the situation becomes more complex when the {\bgps}~are followed by other relational operators, and then connected by \joinopr.
Specifically, {if the \bgps~are followed by \project, it does not affect join elimination because the \project~ can be reordered with the \joinopr~ without impacting the query results.}
% if the {\bgps}~are followed by \project, which will not affect the query results, the \join~can be safely eliminated.
If a \select~follows, employing the \filterrule~allows us to push the filter into the {\bgps}~and then eliminate the \joinopr.
Conversely, if the {\bgps}~are followed by relational operators like \group, \order, or \limit, which alter the results of the {pattern matching},
%~on which these relational operators are applied,
the subsequent \joinopr~connecting the {\bgps}~cannot be eliminated.
In \reffig{rbo}(a), the two \bgps~ can be merged into a single \bgp~ with $v_1$ and $v_3$ as the common vertices, as shown in \reffig{rbo}(b).}

\stitle{\commonrule.}
{For two {\bgps}~connected by binary operators such as \union, \joinopr, and \difference, we design the \commonrule~to identify common subpatterns in the two \bgps, and save the computation cost by matching the common subpattern only once.}
\eat{This is another graph-relational rule designed to reduce computation costs by sharing common query parts.
This approach is especially advantageous when multiple patterns need to be matched, and some of them have overlapping subpatterns.}
For example, when querying the following \cgp:
\begin{lstlisting}[frame=none]
(v1:Person)-[]->(v2:Person)-[]->(:Product)
UNION
(v1:Person)-[]->(v2:Person)-[]->(:Place),
\end{lstlisting}
\commonrule~identifies the common subpattern \code{(v1:Person)-}[] \code{->(v2:Person)} across the binary \code{UNION} operator. It computes and clones the results, which are then expanded for \code{(v2:Person)} -[]\code{->(:Product)} and \code{(v2:Person)-}[]\code{->(:Place)} separately. Finally, the results are combined using the \code{UNION} operation.

\subsection{Type Inference and Validation}
\label{sec:typeinfer}

\begin{algorithm}[t]
    \SetVline
    \SetFuncSty{textsf}
    \SetArgSty{textsf}
    \footnotesize
    \caption{Type Inference and Validation}
    \label{alg:typeinfer}
    \Input{Pattern $P$, Graph Schema $S$}
    \Output{Pattern $P'$ with validated type constraints or INVALID}
    \State{Initialize $Q=\{u | u\in V_P\}$ and sort $Q$ by ascending $|\type_P(u)|$}
    \While{$Q \neq \emptyset$} {
        \State{$u = Q.pop()$}
        \State{$\candi = \emptyset$ and $\candie = \emptyset$}
        \ForEach{$t\in\type_P(u)$}{
            \If{$|\nbr{P}(u)|>0$ \textbf{and} $|\nbr{S}(t)|=0$}{
                \State{remove $t$ from $\type_P(u)$}
            } \Else{
                \State{$\candi = \candi \cup \nbr{S}(t)$} % infer type of edge
                \State{$\candie = \candie \cup \nbre{S}(t)$} % infer type of edge
            }
        }
        \ForEach{$v\in \nbr{P}(u)$}{
            \State{$\type_P(v) = \type_P(v) \cap \candi$} % infer type of adj vertex from edge
            \State{$\type_P(e_{u,v}) = \type_P(e_{u,v}) \cap \candie$} % infer type of edge
            \If{$\type_P(v) = \emptyset$ \textbf{or} $\type_P(e_{u,v}) = \emptyset$}{ % check type
                \Return{INVALID}
            }
            \If{$\type_P(v)$ is updated \textbf{and} $v\notin Q$}{
                \State{insert $v$ into $Q$}
            }
        }
    }
    \State{\textbf{Return} $P'$ with validated type constrains}
\end{algorithm}

{%After applying the heuristic rules, we move on to the optimization of matching \bgps.
%{For clarity and understanding, we simplify the \bgp~ in the logical plan as a graph, where \getvertex~corresponds to vertices, \expandedge~corresponds to edges, and \expandpath~with hop number $l$ corresponds to $l$ consecutive edges.}
As we discussed, \bgps~ in \cgps~ often contain arbitrary type constraints in real applications, which can be \btype, \utype, or \unionall.}
\eat{
Next we show how \gopt~infers and validates the type constraints in \pattern.
As we introduced, the type constraints can be \btype, \utype, or \unionall.
}
{While constraints like \unionall~enhance query flexibility by allowing patterns to be expressed without specific type constraints, they can significantly degrade performance if not properly optimized.
For instance, consider the pattern in \reffig{rbo}(b), where both $v_1$ and $v_2$ are typed with \unionall. In this scenario, the backend engine may have to scan all vertices, as they all meet the type constraint. However, if certain type combinations do not exist in the underlying data, much of this computation is wasted.
As an example, suppose $v_1$ represents a \code{Person}, and $v_2$ is expanded to represent a \code{Place}. If there is no edge connecting $v_2$ to a known \code{Place} $v_3$, the process generates unnecessary intermediate results, exacerbating inefficiency. Furthermore, during {CBO}, imprecise type constraints can lead to significant deviations between estimated and actual costs, resulting in suboptimal query plans.
These issues highlight the critical need for type inference in \bgps.
}
%Even with query optimization,
%the estimated costs can deviate substantially from the actual value due to the imprecise type constraints,
%which may further lead to suboptimal query plans.
%This raises the need for type inference for \bgps.

{For clarity, we first assume a schema-strict context~\cite{Angles_2023}, where the types of vertices and edges are predefined and can be directly utilized for type inference.} A naive solution is to unfold each \utype~and \unionall,
enumerating all possible type combinations including \btypes~only in the pattern, and then validate each combination according to the schema.
However, it is not practical as such combinations can be exponential to the \bgps. %to the number of vertices and edges with \utype~(or \unionall) in \pattern.
% The Pathfinder algorithm \cite{DBLP:conf/grades/SteerALCVV17} infers type constraints for a path pattern using hints from the schema, but is not applicable to general patterns. Moreover, in Pathfinder, the inferred types must be \btypes, leading to potentially enormous inferred patterns similar to the naive solution.
{The Pathfinder algorithm \cite{DBLP:conf/grades/SteerALCVV17} infers type constraints for a path pattern using schema hints, but is not applicable to general patterns. Moreover, Pathfinder requires inferred types to be \btypes, leading to potentially enormous inferred patterns similar to the naive solution}.
% {
%     \cite{DBLP:conf/grades/SteerALCVV17,DBLP:journals/corr/abs-2403-01863} infers type constraints for a path pattern using schema hints, and outputs inferred types to be \btypes~only, leading to potentially enormous inferred patterns similar to the naive solution.
% }
%{In relational databases, type inference is not studied because the types of variables in relational queries are explicitly specified by the tables to which they belong,
%which presents additional challenge as there are no reference methods from relational databases.}
%{\cite{DBLP:conf/grades/SteerALCVV17} proposed the Pathfinder algorithm to infer type constraints for a given path pattern using hints from the schema, but it is not applicable to general patterns. Moreover, the inferred types must be \btypes, resulting in potentially enormous inferred patterns, similar to the output of the naive solution.}

To address the drawbacks, we proposed a novel algorithm in \refalg{typeinfer} to infer types with schema hints,
while preserve the flexibility by \utype~if the inferred results contains multiple types.
\eat{
The basic idea is to iteratively infer and validate type constraints of vertices and edges in the pattern based on the pattern structure and user specified type constraints, as well as type candidates from the graph schema.
The algorithm is presented in \refalg{typeinfer}.
This process continues repeatedly, refining the inferred type constraints until they stabilize and no further changes occur.
During the process, once no valid type constraints can be assigned to a vertex or edge, the algorithm terminates and returns an INVALID flag, indicating that the query pattern with the specified type constraints is not valid and cannot match any subgraphs in the data graph.
}
We start by initializing a priority queue $Q$ to preserve all the vertices in the pattern, %to be refined,
sorted in ascending order of $|\type(u)|$ to ensure that vertices with more specific type constraints are addressed first. % allowing for a more targeted and efficient inference process.
%Then we iteratively refine the type constraints for each vertex $u$ in $Q$ and its adjacent vertices and edges.
%
% \noindent(1) Type Refinement for Vertex $u$ ({lines}~5-10).
% First we assess each type $t$ in $\type(u)$.
% If $u$ has out-neighbors in pattern $P$, but $t$ lacks out-neighbors in schema $S$,
% $t$ cannot be valid and can be pruned. %since no more type constraints can be assigned for $u$'s out-neighbors regarding $\type_b$,
% Otherwise, we update candidates for $u$'s out adjacencies by adding the types of $t$'s out adjacencies in $S$.
%
% \noindent(2) Type Refinement for $u$'s Adjacencies ({lines}~11-17).
% We refine the type constraints for $u$'s out neighbors and edges by intersecting their type constraints with $\candi$ and $\candie$ respectively, ensuring that they align with the constraints ascertained from $u$.
% %Note that this procedure is sensitive to the direction, meaning that out-going (in-coming resp.) adjacencies of $u$ must correspond to the out-going (in-coming resp.) candidate types.
% An empty intersection result implies a failure to assign any valid types, and the algorithm terminates with INVALID, indicating the pattern fails in type checking.
% Otherwise, if the neighbor's type constraint is updated and it is not in $Q$,
% it will be added to $Q$ for further processing.
%
For vertices in the pattern, the algorithm iteratively updates the vertex's type constraint and those of its {neighboring edges and vertices} based on the graph schema connectivity.
If no valid type constraints can be assigned, the algorithm returns an INVALID flag; 
Otherwise, the algorithm terminates when $Q$ is empty, indicating all inferred type constraints are stable, and returns the pattern with these validated constraints. Note that for simplicity and due to space limits, the algorithm only considers outgoing adjacencies, but incoming adjacencies can be handled similarly.

\stitle{Complexity Analysis.}
% {
% %We store pattern $P$ and schema $S$ in adjacency lists.
% Assume the algorithm converges in $k$ iterations.
% In each iteration, to refine $u$'s type constraints, the time complexity is $O(d_S|V_S|)$, where $d_S$ is the maximum vertex degree in $S$.
% To refine for $u$'s adjacencies, it is $O(|V_P||V_S|)$.
% Thus, the overall complexity is $O(k|V_S|(d_S + |V_P|))$.
% In practice, the algorithm converges quickly by starting with the most specific type constraints.
% }
{Assume the algorithm converges in $k$ iterations, the overall complexity is $O(k|V_S|(d_S + |V_P|))$, where $d_S$ is the maximum vertex degree in $S$. In practice, the algorithm converges quickly by starting with the most specific type constraints.}
% {Assume the algorithm converges in $k$ iterations.
% The time complexities of refining $u$'s type constraints and $u$'s adjacencies are $O(d_S|V_S|)$ and $O(|V_P||V_S|)$, respectively, where $d_S$ is the maximum vertex degree in $S$.
% Thus, the overall complexity is $O(k|V_S|(d_S + |V_P|))$.
% In practice, the algorithm converges quickly by starting with the most specific type constraints.}

\begin{remark}
\label{rem:schema-loose}
{We address cases involving schema-loose systems, such as Neo4j, where incomplete schema information prevents the filtering of invalid type constraint assignments in the pattern.
To overcome this, we can retrieve the schema information directly from the data graph during the initialization phase. Schema extraction can be performed using methods such as those described in \cite{DBLP:journals/pvldb/BonifatiDMGJLP22} or system-provided tools like APOC~\cite{cypher_apoc} in Neo4j. Once the schema is established, we can incrementally track changes caused by data modification operations, ensuring the schema remains up to date for subsequent optimization.
If the data is not initially present, optimization is unnecessary at this stage, and the schema extraction process can be deferred until a sufficient amount of data is available.
}
\end{remark}
\vspace*{-1ex}

\begin{figure}[t]
    \centering
    \includegraphics[width=\linewidth]{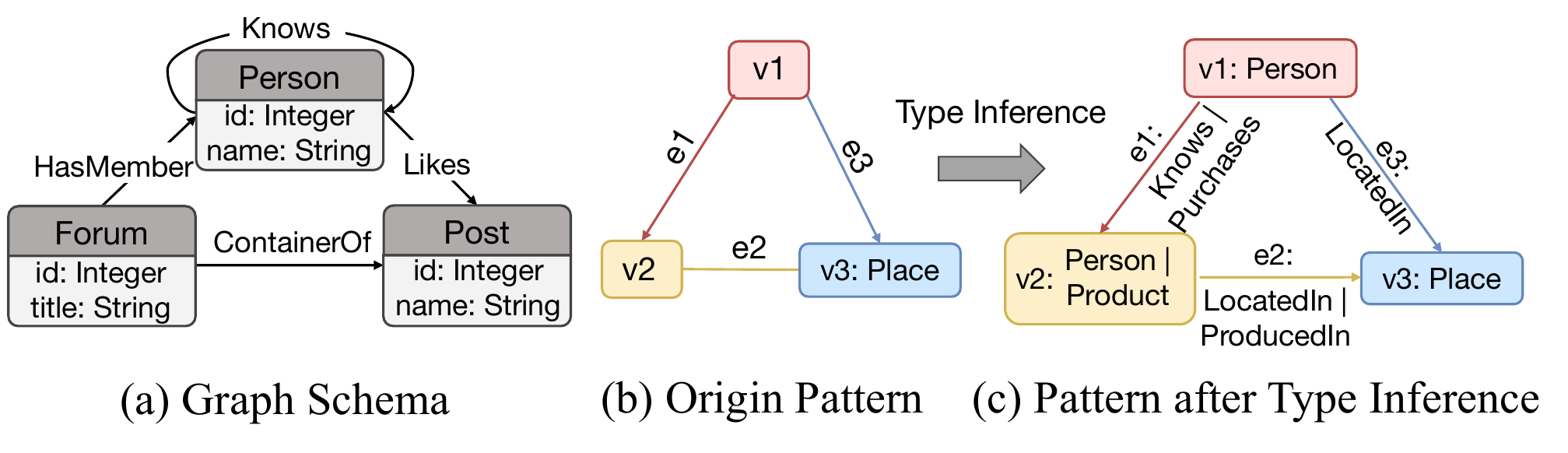}
    \caption{An Example of Type Inference and Validation}
    \label{fig:typeinfer}
    \vspace{-1ex}
  \end{figure}

  \begin{example}
    Given the graph schema in \reffig{typeinfer}(a) and the origin pattern in \reffig{typeinfer}(b),
    the result pattern after type inference is shown in \reffig{typeinfer}(c).
    We start with vertex $v_3$.
    By examining each basic type in $\type(v_3)$, {i.e., \code{Place}},
    %which in this case is \code{Place} and is valid,
    \candi~for $v_3$'s in neighbors is updated as \code{\{Person, Product\}} through edges \code{Person-LocatedIn->Place} and \code{Product-ProducedIn->Place} in schema.
    Thus, $\type(v_1)$ and $\type(v_2)$ are updated to \code{\{Person,Product\}}.
    Then for $v_2$, we update \candi~for $v_2$'s in neighbors through \code{Person-Knows->Person}, \code{Person-Purchase->Product},
    which further narrows $\type(v_1)$ down to \code{\{Person\}}.
    Type constraints of edges can be inferred similarly.
    % After inference, the implicit type constraints are identified as shown in \reffig{typeinfer}(b),
    % which shows the effectiveness of the type inference approach.
\end{example}

\subsection{Cost-based Optimization}
\label{sec:cbo}
%As RBO relies only on predefined rules and ignores data statistics,
We introduce Cost-Based Optimizations (CBO) for processing patterns in \cgps, incorporating data statistics, backend engine variations, and complex type constraints. 
Inspired by prior works\cite{yang2021huge,ammar2018distributed,glogs}, we focus on two principles: (1), high-order statistics~\cite{glogs} for precise cardinality estimation, and (2) hybrid pattern joins combining \wcoj~\cite{ngo2018worst} and binary joins for improved performance.
However, directly applying these techniques face challenges: (1) existing high-order statistics only support patterns with \btypes, and cannot handle \utypes, and (2) backend engines may lack efficient implementations of join operations, particularly \wcoj~(e.g. Neo4j).
We explain how \gopt addresses these challenges below.

%To address these limitations, we introduce \physicalbuilder, enabling backends to register custom physical operators with cost models, and propose a new cardinality-estimation method to support arbitrary type constraints in \bgps, expanding the applicability of existing optimization techniques.}

\subsubsection{Cardinality Estimation}
\label{sec:cardinality}

% Basically, we adopt \glogue~\cite{glogs} as a high-order statistics provider, which provides two query interfaces for a \bgp~$P$, when it contains only \btypes:
% \begin{itemize}
%     \item $\glogvertex(P)$ {returns the precomputed frequency of $P$};
%     \item $\glogedge(P)$ {returns all the preserved edges connected to $P$}, indicating the pattern transformations according to \joinrule~with two types of \join~and \expand.
% \end{itemize}
% However, it does not maintain the patterns with arbitrary type constraints. To address this limitation, we propose a novel cardinality estimation method for arbitrary patterns.
%
{
%We utilize \glogue~\cite{glogs} as a higher-order statistics provider, which precomputes the frequency of \bpatterns~in the data graph to enhance cardinality accuracy.
%However, \glogue~does not support patterns with arbitrary type constraints. To overcome this limitation, we propose a novel cardinality estimation method.
We leverage \glogue~\cite{glogs}, which precomputed the frequencies of \bgps~ as higher-order statistics for cardinality estimation. Techniques from~\cite{glogs}, such as graph sparsification, pattern encoding, and efficient maintenance and lookup in \glogue, are directly utilized. Our focus is on processing patterns with arbitrary type constraints.
}

Basically, given two \eat{induced}subgraphs $P_{s_1}$, $P_{s_2}$ of $P_t$ with $E_{P_t}=E_{P_{s_1}}\cup E_{P_{s_2}}$,
by assuming the independent presence of patterns in the data graph, we compute the frequency of $P_t$ as follows:
\begin{equation}
    \label{eq:joinfreq}
    \freq_{P_t} = \frac{\freq_{P_{s_1}} \times \freq_{P_{s_2}}}{\freq_{P_{s_1}\cap P_{s_2}}}
\end{equation}
where $P_{s_1}\cap P_{s_2}$ denotes the common parts of $P_{s_1}$ and $P_{s_2}$.

{For the case $V_{P_t} \setminus V_{P_s} = \{v\}$, that the edges ${e_1, \ldots, e_n}\in E_{P_t}\setminus E_{P_{s_1}}$ are iteratively appended on $P_s$} and generates intermediate patterns ${P_1, \ldots, P_n}$, where $E_{P_i}\setminus E_{P_{i-1}}=\{e_i\}$ and $P_n=P_t$.
We define the \textit{expand ratio} $\sigma_{e_i}$, to reflect the change in pattern frequency when expanding edge $e_i=(v_i,v)$, as follows: %, which generates $P_i$,
\begin{equation}
    \vspace*{-1mm}
    % \label{eq:expandratio}
    \sigma_{e_i} = \begin{cases}
    \frac{\sum_{t_{e_i}\in \type(e_i)}\freq_{t_{e_i}}}{\sum_{t_{v_i}\in \type(v_i)}\freq_{t_{v_i}}} & \text{if v }\notin V(P_i)\\
    \frac{\sum_{t_{e_i}\in \type(e_i)}\freq_{t_{e_i}}}{\sum_{t_{v_i}\in \type(v_i)}\freq_{t_{v_i}}\times\sum_{t_{v}\in \type(v)}\freq_{t_{v}}} & \text{if v}\in V(P_i)
    \end{cases}
    \nonumber
    \vspace*{-1mm}
\end{equation}
{It should be noted that the type constraints here, $\type(v)$ and $\type(e)$, can be \btype, \utype, or \unionall.}
%$t_{e_i}$ denotes the edge types in $\type(e_i)$, %when it is a \utype, or simply $\type_{e}$ if it is a \btype.
%and $t_{v_i}$, $t_{v}$ refer to the vertex types in $\type(v_i)$, $\type(v)$, respectively.
%Intuitively, $\sigma_{e}$ reflects the change in pattern frequencies when expanding edge $e$. %, by assuming the independent presence of the pattern and expanding edges in the data graph.
Based on this, the frequency of $P_t$ can be estimated as follows:
\begin{equation}
    \vspace*{-1mm}
    \label{eq:expandfreq}
    \freq_{P_t} = \freq_{P_s} \times \prod_{e_i\in E_{P_t}\setminus E_{P_s}}{\sigma_{e_i}}
    \vspace*{-1mm}
\end{equation}
This equation provides a cardinality estimation approach for patterns with arbitrary type constraints based on the subgraph frequencies and the \textit{expand ratio} of the expanded edges.

{Lastly, we {design} a \gloguequery, providing a unified interface \getcount~for the cardinality estimation of arbitrary patterns $P$, as follows:}
Given a \bgp~$P$, if it only contains \btypes, the frequency can be queried from \glogue~directly.
Otherwise, we alternately and iteratively apply \refeq{joinfreq} and \refeq{expandfreq} to estimate $P$'s frequency.
This process continues until the subpattern either exists in \glogue~or \gloguequery(which has been computed and cached), or simplifies to a single vertex or edge whose frequency can be obtained by summing up the frequencies of the contained basic types.

\begin{figure}[t]
    \centering
    \includegraphics[width=0.885\linewidth]{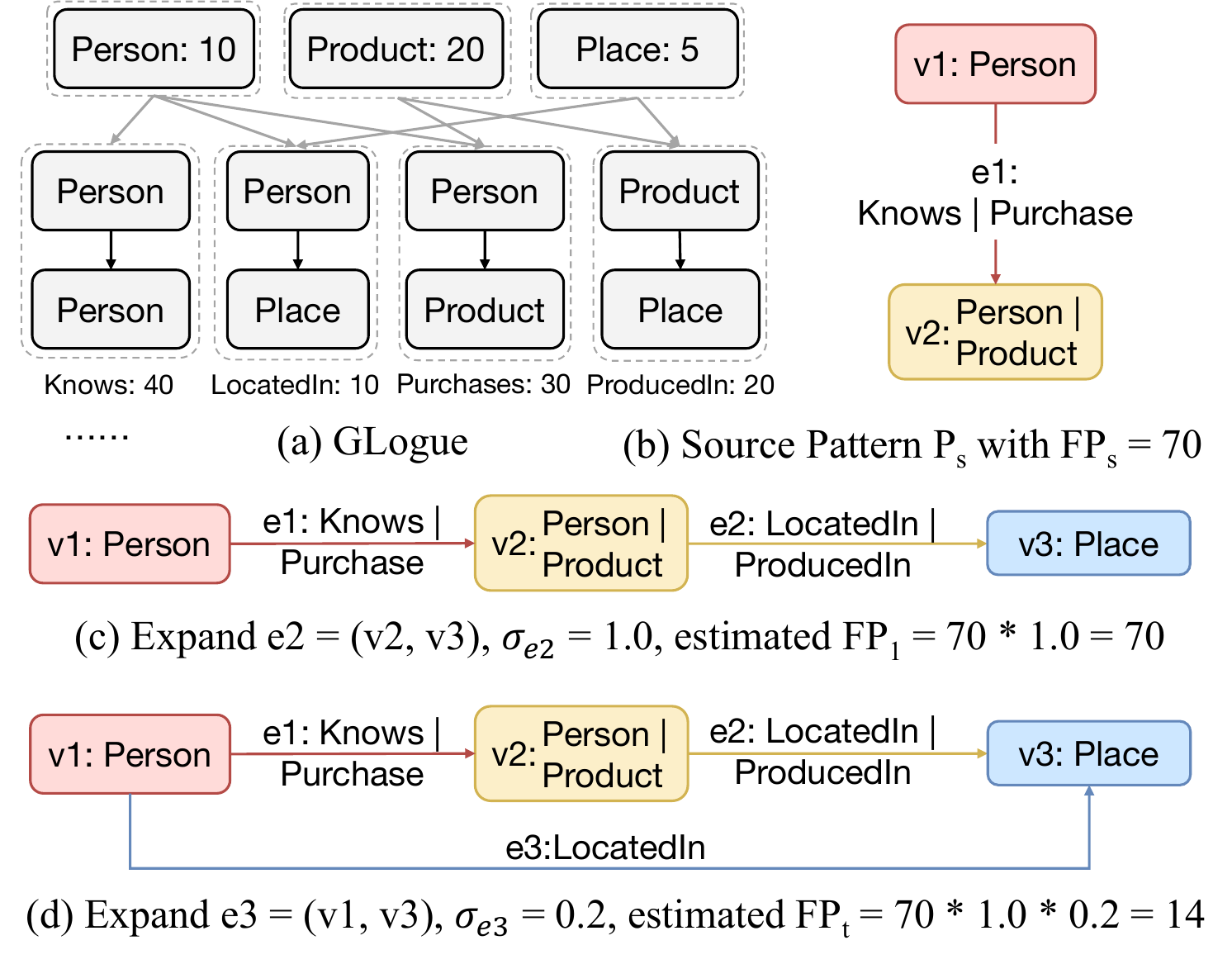}
   % \caption{An Example of Vertex Expansion and Cardinality Estimation}
   \vspace*{-2mm}
    \caption{An Example of Cardinality Estimation}
    \label{fig:cardinality}
\end{figure}

\begin{example}
    {
    Given the \glogue~in \reffig{cardinality}(a)
    and the source pattern $P_s$ in \reffig{cardinality}(b) with $\freq_{P_s} = 70$,
    we show an example of cardinality estimation for pattern $P_t$ in \reffig{cardinality}(c-d).
    First we expand $e_2$ of \utype, as shown in \reffig{cardinality}(c),
    and get
    $\sigma_{e_2} = \frac{\freq_{\code{LocatedIn}}+\freq_{\code{ProducedIn}}}{\freq_{\code{Person}}+\freq_{\code{Product}}} = 1.0$.
    Then we consider $e_3$ of \btype, as shown in \reffig{cardinality}(d),
    and compute $\sigma_{e_3} = \frac{\freq_{\code{LocatedIn}}}{\freq_{\code{Person}}\times\freq_{\code{Place}}} = 0.2$.
    Thus, we have $\freq_{P_t} = 70 \times 1.0 \times 0.2 = 14$}.
    %can estimate the frequency of $P_t$ as
    %$\freq(P_t) = 70 \times 1.0 \times 0.2 = 14$.
\end{example}

\subsubsection{Registerable Physical Plan.}
\label{sec:hybridjoin}
%We incorporate hybrid pattern join implementations in \gopt.
The \joinrule equivalent rule ensures the correctness of pattern transformations during CBO.

\stitle{\joinrule.}
Given data graph $G$ and pattern $P_t$, with $P_{s_1}$ and $P_{s_2}$ where $P_t = P_{s_1} \bowtie_k P_{s_2}$ and $\bowtie_{k}$ is the join operator with join key $k = V_{P_{s_1}} \cap V_{P_{s_2}}$.
Let $R(P,G)$, or $R(P)$ for brevity, represent the {results of matching } $P$ in $G$.
{Under homomorphism-based matching semantics (with an all-distinction operator for other semantics, as discussed in \refrem{semantics})}, {$R(P_t)$} can be computed by:
\begin{equation}
R(P_t) = R(P_{s_1}) \bowtie_{k} R(P_{s_2})
\end{equation}

\stitle{{\physicalbuilder.}}
{
    Following existing works~\cite{mhedhbi2019optimizing, yang2021huge}, two execution strategies, binary join and vertex expansion, are commonly used to implement \joinrule.
    However, different backends may implement these strategies differently, potentially leading to suboptimal query plans if costs are not accurately calculated. To address this, we introduce a \physicalbuilder~interface, allowing backends to register their custom physical operators and corresponding cost models:
% that enables backends to register their custom physical operators along with their corresponding cost models:
}
\begin{lstlisting}
interface PhysicalSpec {
    // Compute the cost of joining two patterns.
    // Recall that GlogueQuery provides cardinality of patterns.
    Double computeCost(GlogueQuery gq, Pattern Ps1, Pattern Ps2);
}
\end{lstlisting}
{
We use Neo4j and \gs~as examples to show registering custom physical operators with their cost models via \physicalbuilder:
}

{
(1) \textbf{Binary Join}: denoted as $\join({P_{s_1}, P_{s_2}} \rightarrow P_t)$, this strategy computes the mappings of  $P_{s_1}$  and  $P_{s_2}$  separately and then joins the results to produce the mappings of $P_t$. Both Neo4j and \gs~ implement this using \hashjoin~ and register the corresponding cost as $\freq_{P_{s_1}} + \freq_{P_{s_2}}$ according to~\cite{glogs} in the \physicalbuilder.
}

{
(2) \textbf{Vertex Expansion}: Denoted as $\expand(P_s \rightarrow P_t)$, this strategy applies when $V_{P_t} \setminus V_{P_s} = \{v\}$, meaning $P_t$  is formed by appending a vertex $v$ to  $P_s$. Neo4j implements this with the \expandinto~operator.
\gs~ utilizes the \expandintersect~ operator, based on algorithms proposed in~\cite{mhedhbi2019optimizing}.
Each system can register its physical operators with their respective cost models, as shown below:
}
\begin{lstlisting}
// Neo4j's registration
class ExpandInto implements PhysicalSpec {
    @Override
    Double computeCost(GlogueQuery gq, Pattern Ps, Pattern Pv) {
        // Pv denotes the subpattern that:
        // Vpv = Vpt \ Vps, Epv = Ept \ Eps
        Pattern Pi = Ps, cost = 0
        for each e in Pv {
            Pi <- append e to Pi
            cost += gq.getFreq(Pi)
        }
        return cost;
    }
}
// GraphScope's registration
class ExpandIntersect implements PhysicalSpec {
    @Override
    Double computeCost(GlogueQuery gq, Pattern Ps, Pattern Pv) {
        return |Pv| * (gq.getFreq(Ps));
    }
}
\end{lstlisting}

\begin{figure}[t]
    \centering
    \includegraphics[width=0.9\linewidth]{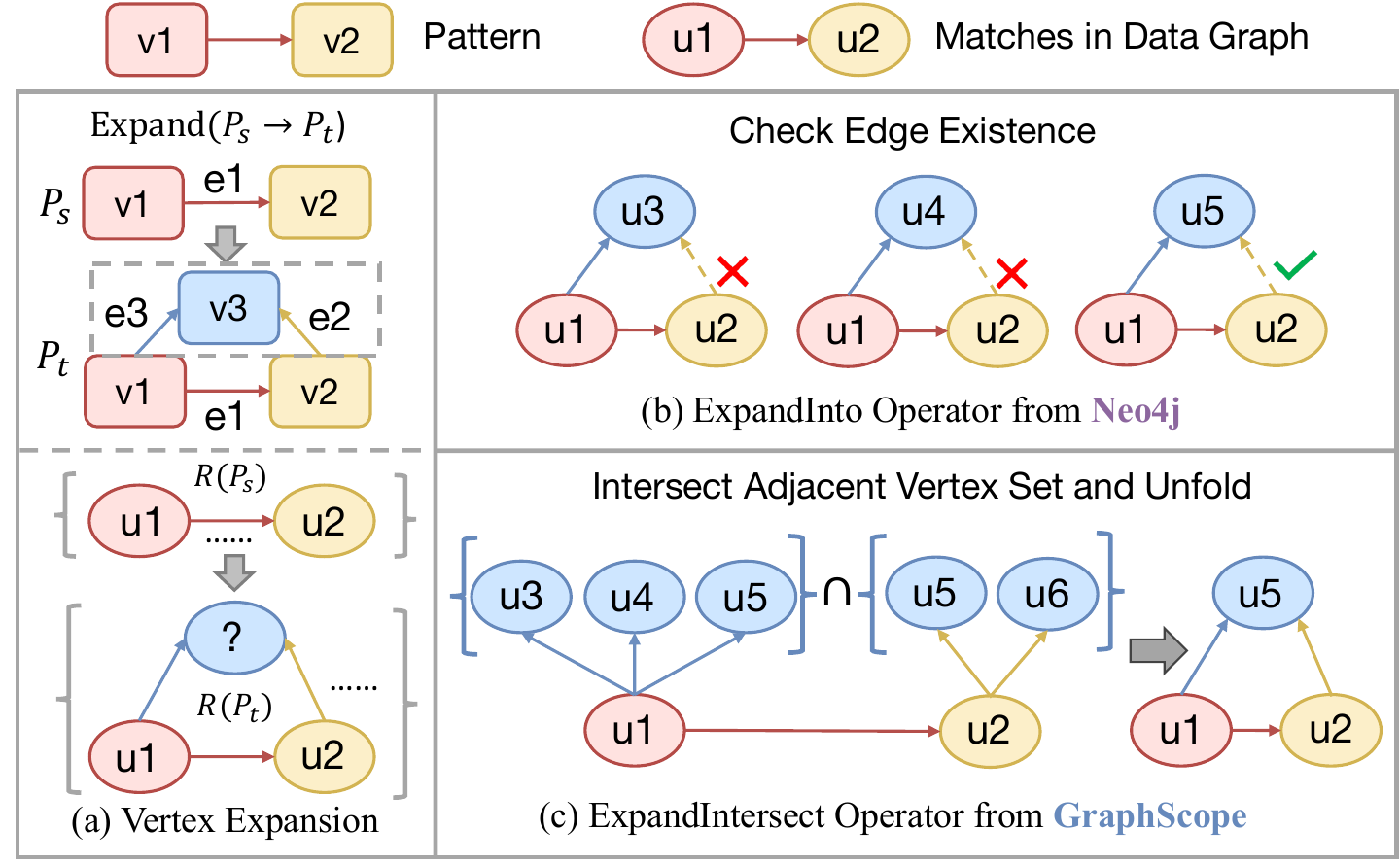}
    \caption{An Example of Vertex Expansion Implementations}
    \label{fig:cbo}
\end{figure}

\example{
    We illustrate the different vertex expansion implementations used by Neo4j and \gs~in \reffig{cbo}.
    Initially, pattern $P_s$ matches to $(u_1, u_2)$ in the data graph (\reffig{cbo}(a)). To match $P_t$ by expanding $e_3$ and $e_2$,
    Neo4j first expands $e_3$, yielding three mappings (\reffig{cbo}(b)), and then performs \expandinto~to find matches for $e_2$ connecting $u_2$ to $u_3$, $u_4$ and $u_5$, respectively. As \expandinto~flattens the intermediate matching results, the cost is the sum of frequencies of intermediate patterns.
    In contrast, \gs~uses \expandintersect, which begins by finding the match set $R(P_1)$ by expanding $e_3$, yielding one intermediate result. It then expands $e_2$ and intersects the matched set with $R(P_1)$ to obtain $R(P_t)$, and finally unfolds the match set after expanding all the edges (\reffig{cbo}(c)). \expandintersect~reduces computation by avoiding flattening intermediate results, with the cost defined in the snippet.
}

\stitle{{Cost Model.}}
In general, we adopt a cost model for {\cgps}~that considers both communication and computation cost.
The communication cost is defined as the number of intermediate results during execution,
due to fetching remote intermediate results for further processing in distributed systems.
The computational cost is defined as the total cost of the operators in the physical plan,
{each by \computecost~method in \physicalbuilder},
adjusted by a normalized factor $\alpha_{op}$
to reflect each operator's relative processing expense.
Specifically, if the integrated engine employs sequential processor, such as Neo4j, the communication cost can be ignored.

\begin{algorithm}[t]
    \SetVline
    \SetFuncSty{textsf}
    \SetArgSty{textsf}
    \footnotesize
    \caption{Graph Optimizer}
    \label{alg:cbo}
    \Input{Pattern $P$, \gloguequery~$GQ$}
    \Output{Optimal $(plan,cost)$ for $P$}
    \State{$(plan^*, cost^*) \leftarrow \init(P, GL)$}
    \State{Initialize map $M$ = \{p:(plan, cost)\} with patterns of size 1 and 2 precomputed}
    \State{RecursiveSearch($P$, $GQ$, $M$, $cost^*$)}
    \State{\textbf{Return} $M.get(P)$}

    \vspace*{0.1cm}
    \textbf{Procedure} RecursiveSearch($P$, $GQ$, $M$, $cost^*$)\\
    \State{\textbf{if} $M$ contains $P$ \textbf{then} {Return}}

    \For{$e\in \getcandi(P)$} {
        \State{\textbf{if} $\lbound(M, e)\ \geq cost^*$ \textbf{then} {continue}}
        \If{$e$ is of $\expand(P_s\rightarrow P)$}{
            \State{RecursiveSearch($P_s$, $GQ$, $M$, $cost^*$)}
            \State{$cost'$=$M$.$\getcost(P_s)$+$GQ$.$\getcount(P)$+$e$.$\computecost$($GQ$,$P_s$,$P_v$)}
        } \ElseIf{$e$ is of $\join(\{P_{s_1}, P_{s_2}\}\rightarrow P)$} {
            \State{RecursiveSearch($P_{s_1}$, $GQ$, $M$, $cost^*$)}
            \State{RecursiveSearch($P_{s_2}$, $GQ$, $M$, $cost^*$)}
            \State{$cost' = M.\getcost(P_{s_1}) + M.\getcost(P_{s_2}) + GQ.\getcount(P) + e.\computecost(GQ, P_{s_1}, P_{s_2})$}
        }
        \If{$cost' < M.\getcost(P)$}{
           \State{update $P$ in $M$ with the new $plan'$ and $cost'$}
        }
    }
\end{algorithm}

\vspace*{-5ex}
\subsubsection{Top-Down Search Framework}
\label{sec:topdown}
Based on new cardinality estimation methods and backend-registered cost models, we introduce a top-down search framework for optimizing \bgps~in \refalg{cbo}. 
We begin with a greedy search to find an initial solution as a tight bound for further exploration.
A map $M$ is then initialized to store optimal plans for the subgraphs of $P$, with costs for single vertices and edges set to their frequencies.
We then search for the optimal plan for $P$ in a top-down framework.
In the recursive search, \getcandi~identifies candidate edges for pattern transformation targeting $P$ by enumerating the subgraphs.
For each candidate, we prune non-optimal search branches based on \lbound. %which will be explained in details later ({lines}~9-10).
After that, we continue searching for the optimal plan for sub-patterns (lines~10, 13-14),
until they have already been computed and stored in $M$ ({line}~6).
Using the subplan's cost, we compute $P$'s cost by accumulating communication costs (if not a single-node backend) and computation cost based on the backend-registered \physicalspec~(line 11, 15).
If $P$'s cost is lower than previous bests,
we update the optimal plan for $P$ in $M$~({lines} 16-17).

\stitle{Initialize with Greedy Search.}
\init~is designed to identify a good initial solution as a tight bound for branch pruning.
Starting with the queried pattern $P$, it iteratively removes candidate edges with the smallest cost, including \expand~and \join, in a greedy manner,
with the edges' costs and subpatterns' frequencies being accumulated to the pattern cost,
until reducing the pattern to a single vertex.
This initial solution helps \refalg{cbo} prune unpromising branches, enhancing overall efficiency.

\stitle{Prune Branch with Lower Bound.}
We design pruning strategies to filter out the search branches that cannot be optimal (line~9-10).
Consider a candidate edge $\expand(P_s\rightarrow P)$.
If $P_s$ has been previously searched and its cost already exceeds $cost^*$, 
or, if the noncumulative cost of $P$ (i.e., $\freq_{P_s}+e.\computecost(P_s,P_v)$) exceeds $cost^*$, the branch can be pruned.
%Note that except \bpatterns~whose frequencies can be retrieved from \glogue, for \upatterns' frequencies, which can be computed according to \refeq{joinfreq} and \refeq{expandfreq} during the process, we cache the results in \glogue~(line~13-14, 19-20) to avoid redundant computation.
%Once meeting a \upattern~whose frequency has already been computed before, we can get it from \glogue~directly for pruning or further computation.
Similar pruning strategies can be applied when the candidate edge $e$ is of $\join(\{P_{s_1}, P_{s_2}\}\rightarrow P)$.

\section{System Implementation}
\label{sec:impl}

% \gopt~is built upon the Apache Calcite~\cite{Begoli_2018} framework, a versatile platform for dynamic data management with a focus on relational databases.
% By leveraging Calcite's \code{RelNode} structure, \gopt integrates graph operators as subclasses and uses the platform's native relational operators.
% This approach allows \gopt~to benefit from Calcite's robust optimization techniques and extends its functionality to handle graph pattern optimization challenges effectively.
% The optimization of both pattern and relational queries in \gopt~is executed using these integrated \code{RelNodes}.

%Within this integration, graph operators are implemented as extensions of Calcite's relational operators using the \code{RelNode} structure, facilitating a cohesive optimization process for both graph patterns and relational queries.
\eat{
\gopt~is built upon the Apache Calcite framework~\cite{Begoli_2018}, a versatile platform for dynamic data management with a focus on relational databases.
By integrating with Calcite, \gopt~provides a unified approach for optimizing \patrels,
to ensure that \gopt~benefits from the well-established relational optimization techniques in Calcite,
while extending its capabilities to address the unique challenges introduced by graph pattern optimization.
}
\gopt~is built upon the Apache Calcite framework \cite{Begoli_2018}, a dynamic data management platform with a focus on relational databases.
By integrating with Calcite, \gopt~provides a unified approach for optimizing \patrels, combining proposed graph optimization techniques with the relational optimization techniques already present in Calcite.
To build \ir, the graph operators are implemented as subclasses of the fundamental \code{RelNode} structure in Calcite to ensure smooth integration, facilitating a cohesive optimization process for both graph patterns and relational operations.

\eat{
\stitle{Intermediate Representation}.
In building the IR operators for \gopt, we utilized the fundamental \code{RelNode} structure from Calcite.
We implemented graph operators as subclasses of \code{RelNode} to ensure smooth integration with the Calcite ecosystem,
and directly leveraged the existing relational operators provided by Calcite.
The optimization process then proceeds on query plans composed of these \code{RelNodes}.
}
\stitle{Rule-based Optimization}.
In \gopt, we employ the HepPlanner, an RBO planner in Calcite, to implement our RBO techniques.
Developing new heuristic rules involves two primary steps:
(1) define a condition that determines the applicability of the rule, and
(2) specify an action that modifies the subplan when the condition is met.
For instance, in \filterrule,
when a \select~operator with filters targeting a tag associated with a graph operator is identified (the condition),
we move the filter expression into the graph operator (the action).
%This modification allows us to directly eliminate unwanted data during the graph data retrieval process.
Furthermore, we incorporate Calcite's existing rules for relational operators into \gopt,
thereby leveraging the established relational optimizations.

% HepPlanner in Calcite is a rule-based optimizer (RBO).
% It applies a set of predefined rules to the logical plan to get a new equivalent one that is assumed to be more efficient in a heuristic way.
% It allows users to register customized rules and apply them in the optimization process.
% To register a rule, two necessary information should be provided:
% (1) a condition to check whether the rule can be applied,
% and (2) the action performed on the sub-plan once it matches the condition.

% We register all the new developed heuristic rules in \gopt~in Calcite.
% For example, for the \filterrule,
% once we find a \select~operator with filter expressions operating on a tag that refers to a graph operator (the condition),
% we push the filter expression into the graph operator (the action),
% with which we can directly filter out the undesired data during graph data access.

\stitle{Cost-based Optimization}.
We further adopt the VolcanoPlanner, a CBO planner in Calcite, to implement the CBO techniques {for patterns}.
First, we introduce a new metadata provider, %\code{GraphMdQuery},
% which provides information such as the computational cost and estimated number of output results associated with each operator.
% We implement metadata handlers for graph operators to provide cost and cardinality estimation
% based on the cost model and cardinality estimation methods outlined in \refsec{cbo}.
% For relational operators, we utilize the built-in metadata handlers provided by Calcite.
{\code{GraphMdQuery}, which provides metadata for both graph and relational operators.
For graph operators, we implement the handlers for \code{RowCount} using \gloguequery~and for \code{NonCumulativeCost} using \physicalbuilder, as detailed in \refsec{cbo},
to deliver more accurate cardinality estimates and backend-specific cost estimations.
For relational operators, we leverage Calcite's built-in metadata handlers.
}
Then to optimize the \bgps, we implement \joinrule~through \eat{\join~and \expand, corresponding to the hybrid join strategy}{\hashjoin, \expandinto(on Neo4j) and \expandintersect(on \gs), to support the hybrid join strategy}.
%Both physical operators are also implemented as subclasses of \code{RelNode} in Calcite, maintaining a consistent representation.
Leveraging these transformation rules, the VolcanoPlanner can produce equivalent plan sets for patterns, where each set includes all possible plans for a specific induced subgraph of the original pattern.
After integrating {\code{GraphMdQuery}} and registering \joinrule~into VolcanoPlanner,
we invoke the CBO process to optimize \patrels~in a top-down manner.
As the search progresses, the VolcanoPlanner consults \code{GraphMdQuery} for estimated cardinalities and costs to calculate the total cost of the plan.
%After determining the search order of the pattern with the lowest cost, the pattern's execution plan
%will be wrapped into a special relational Scan operator.   if needed, the optimizer moves on to optimize the relational part of the query.}
% of \patrel~based on \code{GraphMdQuery},
%as it also provides the metadata for relational operators.
% Note that \code{GraphMdQuery} in \gopt~serves as a comprehensive metadata provider for both graph and relational operators,
% facilitating a unified CBO process across the entire query plan.
% While currently a two-phase CBO is devised for \gopt,
% thus the system is flexible enough to execute a single-phase CBO on the full query plan,
% by registering all the rules for equivalent transformations for both graph and relational parts at the beginning,
% when it proves advantageous for optimization purposes. We will also explore this in future work.
% \todo{two phase or single phase?}
%Note that we can also execute a single-phase CBO on the full query plan by registering all the equivalent transformation rules for both graph and relational parts at the beginning, when it proves advantageous for optimization purposes. We will also explore this in future work.

\begin{remark}
    \label{rem:impl_calcite}
    {
    While optimizing \cgps, we encapsulate \bgps~ within a specialized relational \scan~ operator, with its cost determined by the \bgp~ frequencies, which enables us to leverage Calcite’s built-in metadata handlers and relational optimization rules to enhance the relational components of ~\cgps. This approach allows \gopt~ to benefit from Calcite’s robust relational optimization techniques. For example, when executing a \group~ operator, Calcite’s \kw{AggregatePushDown} rule can help reduce intermediate results and improve efficiency.
    }

    {
        Moreover, more sophisticated cost estimation is required for \bgps~ with filters after they are incorporated into \bgps~ (as described in \filterrule~ in \refsec{rbo}). Currently, we followed~\cite{glogs} to predefine selectivity values (e.g., 0.1) for vertices and edges with filtering conditions.
        In future work, we aim to integrate advanced selectivity estimation techniques, such as histograms and sampling provided by Calcite, to further improve optimizing \bgps~ with filters.
    }
\end{remark}

\stitle{Output Format}.
We use the Google Protocol Buffers (protobuf)~\cite{protobuf} to define the format of the physical plan output by \gopt. Protobuf provides a platform-independent format, enabling seamless integration of \gopt~with various backends.
%We have meticulously crafted the protobuf definitions %for the physical operators and the overall structural layout of the physical plan.
%We have also developed a protobuf serializer within \gopt~to facilitate the translation of the optimized query plan into a protobuf format.
%The use of protobuf offers a significant advantage due to its platform-independent nature, enabling \gopt~to be effortlessly integrated with a myriad of backend engines.
Specifically, we have integrated \gopt~into Alibaba's \gs~platform, allowing the optimized physical plans to be executed on the distributed dataflow engine \gaia~\cite{qian2021gaia} through a \code{PhysicalConverter}. This integration enhances the efficiency and scalability of handling \patrels~in \gs.
Similarly, we have also integrated \gopt~into Neo4j to demonstrate its compatibility and effectiveness in various graph systems.

% VolcanoPlanner in Calcite is a cost-based optimizer (CBO).
% The basic idea is that it transforms the logical plan into a set of equivalent logical plans,
% estimates the cost of each plan based on metadata (achieved by the interface `RelMetadataQuery'),
% and finally returns the plan with the lowest cost.
% It allows users to provide:
% (1) the `RelMetadataQuery' provides APIs for the metadata querying,
% including but not limited to the estimated row count output by the operator,
% the noncumulative estimated cost of the operator, and the cumulative cost of the plan.
% Users can register customized `RelMetadataQuery' implementations by providing the metadata query handler as a plugin,
% e.g., a new handler to estimate the row count.
% (2) the same as in the RBO, the transformation rule that can be applied to the plan to generate a new equivalent plan.
% Besides, it provides the search manner of both bottom up and top down, switched by a flag.
% The top-down search paradigm in Calcite combines pruning strategies that
% if the cost of a sub-plan is higher than the current lowest cost, it will be pruned.

% We incorporate the \gopt~into Calcite by plugin a new metadata provider:

% The metadata query handler is responsible for answering the metadata query from Calcite.

\vspace*{-1mm}
\section{Experiments}
\label{sec:exp}
{We conducted experiments to demonstrate the effectiveness of the proposed optimization techniques in \gopt~ and its ability to integrate with existing systems. These experiments showcase how \gopt~ readily enables support for multiple query languages while enhancing query processing performance.
%So far, we have integrated \gopt~ into \gs\cite{graphscope} and Neo4j~\cite{neo4j}.
}

\vspace{-0.5em}
\subsection{Experiment Settings}
\vspace{-0.5em}
\stitle{Benchmarks.}
{
We utilized the Linked Data Benchmark Council (LDBC) social network benchmark~\cite{ldbc_snb} to test \gopt.
As shown in \reftab{datasets}, $4$ datasets were generated using the official data generator, where $G_{sf}$ denotes the graph with scale factor $sf$. {As our focus is on optimization techniques, we used the initial snapshot of each dataset and did not account for dynamic updates.}
We adopted the LDBC Interactive workloads $IC_{1\ldots 12}$, and Business Intelligence workloads $BI_{1\ldots 14, 16, 17, 18}$, excluding $IC_{13, 14}$ and $BI_{15,19,20}$.
These excluded queries either replies on built-in shortest-path
algorithm or stored procedures, and are not currently supported in \gs.
In addition, we designed three query sets: $QR_{1\ldots 8}$ to evaluate heuristic rules, $QT_{1\ldots 5}$ to test type inference, and $QC_{1\ldots 4(a|b)}$ to assess CBO. {All the queries can be found in~\cite{bench_queries}. By default, these queries were implemented in Cypher, with LDBC workloads using the official Cypher implementations~\cite{ldbc_snb_impl}. Certain queries written in Gremlin were manually translated from their Cypher counterparts.}
}

\begin{table}
  \small
  \centering
  %\vspace*{-0.5em}
  \caption{The LDBC datasets.}
  \vspace*{-0.5em}
  \label{tab:datasets}
  \begin{tabular}{ l | r | r | r  }
  \hline
    \bf{Graph} & \bf{$|V|$} & \bf{$|E|$} & \bf{Size} \\ %& \bf{Time} (min) \\
    \hline
    $G_{30}$ & $89$M & $541$M & $40$GB \\ %& $10$ \\
    \hline
    $G_{100}$ & $283$M & $1,754$M & $156$GB \\ % & $34$ \\
    \hline
    $G_{300}$ & $817$M & $5,269$M & $597$GB \\ %& $101$ \\
    \hline
    $G_{1000}$ & $2,687$M & $17,789$M & $1,960$GB \\ %& $353$ \\
    \hline
    \end{tabular}
  \end{table}

  \stitle{Compared Systems.}
  {
    We conducted experiments on two systems:\\
  (1) \gs~v0.29.0~\cite{graphscope_github}, a distributed graph system supporting Gremlin queries. It optimizes queries using a rule-based optimizer and executes them on the distributed dataflow engine \gaia~\cite{qian2021gaia}.\\
  (2) Neo4j~v4.4.9~\cite{neo4j}, a widely used graph database supporting Cypher queries. It optimizes queries using its \cypherplanner, which employs heuristic rules and cost-based optimization, and runs on a single machine using Neo4j's interpreted runtime.\\
  %Gremlin parser and rule-based optimizer in \gs, and the Cypher parser and \cypherplanner~in Neo4j.
  For each query, we highlight the following three execution plans:
  \begin{itemize}[noitemsep,topsep=0pt]
  \item \gopt-plan: The optimized plan generated by \gopt. Note that the \gopt-plans for running on Neo4j and \gs~ may be different due to the \physicalbuilder~ (\refsec{cbo}).
  \item Neo4j-plan: The optimized plans generated by Neo4j's optimizer \cypherplanner. Although Neo4j's execution engine may not be the most competitive, its optimizer remains as a strong baseline, being one of the most well-developed in the community.
  %\item GS-plan: The optimized plans generated by \gs's rule-based optimizer.
  \end{itemize}
  %This integration allows the optimized plans generated by \gopt, termed as \gopt-plans, to be executed seamlessly on both systems.
  %For comparison, we also evaluate the performance using the optimized plans generated by Neo4j's native \cypherplanner~(referred to as Neo4j-plans) and the rule-based optimizer in \gs~(referred to as \gs-plans).
  %Note that we cannot directly execute Neo4j-plans on \gs, as \cypherplanner~does not provide interfaces for integration with other execution engines. Therefore, we manually translated Neo4j-plans into the protobuf format used by \gs~to enable the comparison.
  %\todo{but we only runs Neo4j-plan on \gs.}
  }

  \stitle{Configurations.}
  We conducted tests on a cluster of up to 16 machines, each with dual 8-core Intel Xeon E5-2620 v4 CPUs at 2.1GHz, 512GB memory, and a 10Gbps network.
  To avoid disk I/O effects, all graph data was stored in memory. For the distributed experiments on \gs, vertices and properties were randomly assigned to machines, while edges and properties were placed on machines hosting their source and destination vertices.
  %We ran our tests on a cluster configured with up to 16 machines, each equipped with dual 8-core Intel Xeon E5-2620 v4 CPUs at 2.1GHz and 512GB of memory, connected by a 10Gbps network.
 % To avoid the effects of disk I/O, all graph data was kept in main memory. {For the distributed experiments on \gs,} we randomly assigned vertices and their properties to different machines, while edges and their properties were placed on the machines hosting their respective source and destination vertices.
  % In our analysis, we excluded the query parsing and optimization time, as it is minimal compared to query runtime.
  We excluded query parsing and optimization time from our analysis due to its minimal impact on runtime.
  Queries exceeding 1 hour were marked as \ot.

  \vspace*{-3mm}
  \subsection{Micro Benchmarks}
  \label{sec:micro_exp}
{
  We evaluated the effectiveness of \gopt’s individual optimization techniques in \refsec{opt}, namely heuristic rules, type inference, and cost-based optimizations. To demonstrate \gopt's capability in optimizing different query languages, we further write queries in Gremlin, and compared the performance of GS-plans with \gopt-plans. The experiments were conducted on $G_{30}$ using a single machine with 32 threads, with the queries executed on \gs.
   %be seamlessly integrated into existing systems and provide immediate benefits, we further compared the performance of GS-plans with \gopt-plans. The experiments were conducted on $G_{30}$ using a single machine with 32 threads, with the queries executed on \gs.
  %Furthermore, we compare the performance by executing the optimized plans, \goptplan~and \gsplan~to highlight the overall improvements brought by \gopt~to \gs.
}
{Other than the “Optimizing Gremlin Queries” test, where all \gopt~optimization techniques were applied, we conducted controlled experiments to isolate the effects of specific techniques. For example, when evaluating heuristic rules, we ensured that the queries included only explicit types and disabled type checker and CBO to prevent interference from other techniques.
}
  % In this subsection, we evaluate the effectiveness of \gopt's optimization techniques, including type inference, heuristic rules, and cost-based optimization, using a small-scale dataset $G_{30}$ on a single machine with 32 threads.
  % LDBC and JOB queries are further applied to assess \gopt's performance with real-life queries, offering insights into its practical capabilities.

  \begin{figure}[tb!]
    \vspace*{-1em}
    \begin{center}
    \end{center}
    \begin{center}
    %%%%%%%%%%%%%%%%% Exp-1: type inference
    %%%%%Fig 1:
    \setlength{\subfigcapskip}{-6bp}
    \subfigure[Heuristic Rules]{\label{fig:micro_exp_rbo}
    {\includegraphics[width=4cm, height=2.8cm]{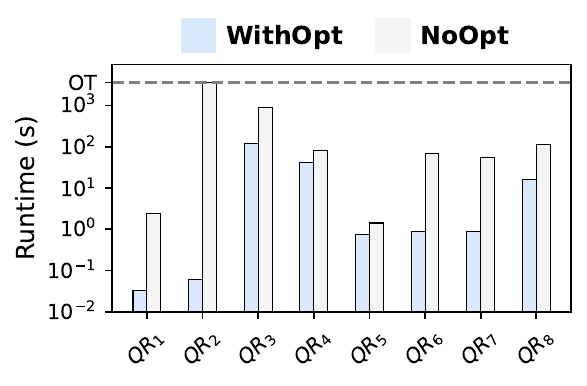}}}
    \vspace{-2ex}
    %%%%%Fig 2:
    \setlength{\subfigcapskip}{-6bp}
    \subfigure[Type Inference]{\label{fig:micro_exp_type_inference}
    {\includegraphics[width=4cm, height=2.8cm]{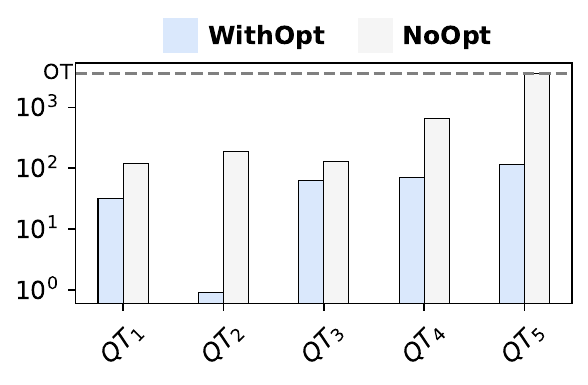}}}
    \vspace*{-3ex}
    %%%%%Fig 3:
    \subfigure[CBO]{\label{fig:micro_exp_cbo}
    {\includegraphics[width=4cm, height=2.8cm]{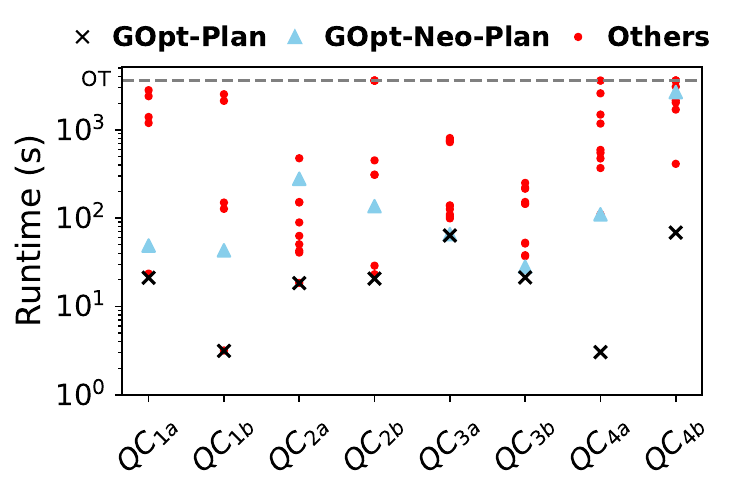}}}
    \subfigure[Cardinality Estimation]{\label{fig:micro_exp_cardinality}
    {\includegraphics[width=4cm, height=2.8cm]{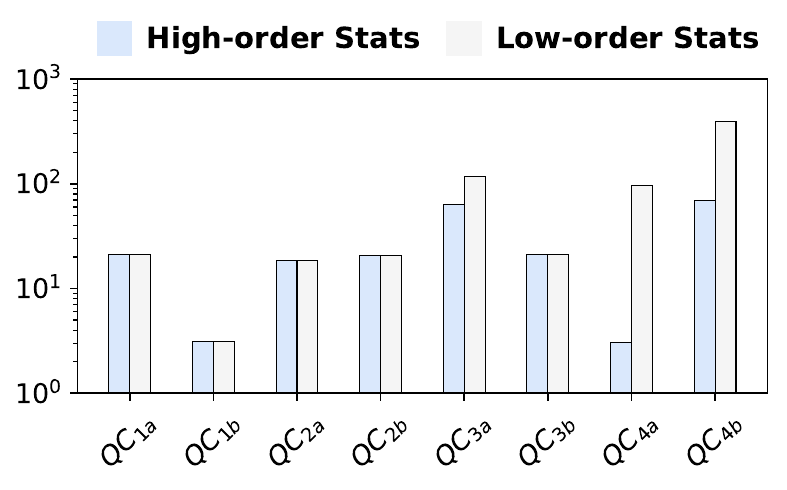}}}
    %%%%%Fig 4:
    \subfigure[Optimizing Gremlin queries on \gs]{\label{fig:micro_exp_e2e}
    {\includegraphics[width=8cm, height=2.8cm]{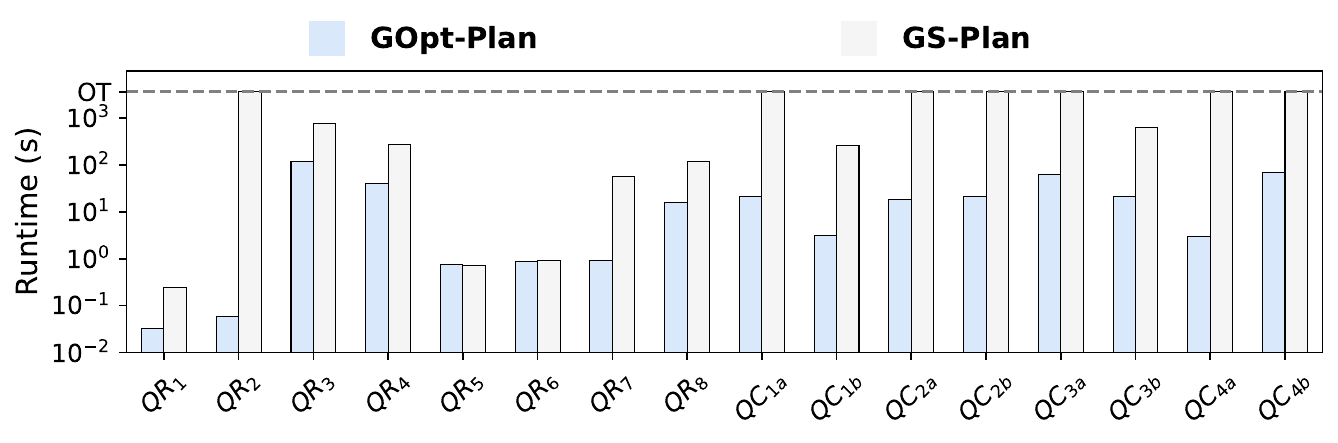}}}
    % \subfigure[\scriptsize{JOB Queries on IMDB}]{\label{fig:micro_exp_job}
    % {\includegraphics[height=2.8cm, width=8cm]{exp/job.pdf}}}
    \vspace*{-0.5em}
    \caption{Results of Micro Benchmarks on \gs.}
    \vspace*{-1ex}
    \label{fig:micro_exp}
    \end{center}
  \end{figure}

  \stitle{Heuristic Rules.}
  {To evaluate the heuristic rules, we compared performance with and without their application. }
 {As shown in \reffig{micro_exp_rbo}, \filterrule~($QR_{1,2}$) demonstrated the most significant impact, improving performance by orders of magnitude. \trimrule~($QR_{3,4}$) achieved an average performance improvement of $4.8\times$, \joinelimrule~($QR_{5,6}$) enhanced performance by $40.2\times$, and the \commonrule~($QR_{7,8}$) achieved a $34.3\times$ improvement.}
  These findings, particularly the substantial gains achieved with the \filterrule, underscore the importance of \ir, which facilitates the effective co-optimization between pattens and relational operators.
  % a unified representation (\ir) of \bgps~and relational operators, facilitating effective co-optimization between them.}

  \stitle{Type Inference.}
  We tested the impact of type inference by comparing query execution performance with it enabled versus disabled {on queries without explicit type constraints}. As illustrated in \reffig{micro_exp_type_inference}, type inference consistently accelerates all queries, with an average speedup of $51\times$.
  Notably, $QT_2$ achieves a $209\times$ improvement, thanks to \gopt's capability to infer type constraints, thereby avoiding unnecessary traversal of irrelevant vertices and edges.

  \stitle{Cost-based Optimization.}
  We tested CBO using $QC_{1\ldots 4(a|b)}$, focusing on pattern matching for a triangle, a square, a 5-path, and a complex pattern consists of 7 vertices and 8 edges.
  {%Heuristic rules were disabled, and to eliminate the effects of type inference, type constraints were explicitly defined,
  Each query has two variants: $a$ for queries with \btypes~and $b$ with \utypes.}
  For performance comparison, we designed up to 12 unique execution plans for each query,
  including: one {\goptplan} (marked with ``$\times$''); one {GOpt-Neo-Plan (marked with ``$\Delta$''), which is generated by \gopt with deliberate computation of the cost of \expandintersect~using Neo4j’s cost of \expandinto}; and up to 10 randomly generated plans (marked with red ``$\circ$'').
  As shown in \reffig{micro_exp_cbo}, the results highlight the vital role of CBO in pattern processing and demonstrate \gopt's effectiveness in utilizing backend-specific cost estimation.
  { The \goptplan~consistently outperformed, averaging $14.4\times$ faster than GOpt-Neo-Plan, thanks to \gopt's \physicalbuilder~enabling backend-specific costs.
  This comparison underscores the importance of using a cost model tailored to the backend system, as mismatched models can result in suboptimal plans.}
  Additionally, \gopt's plan was $117.8\times$ faster the average of the randomized plans.
  While \gopt~may not always achieve the absolute best performance due to its reliance on estimated costs, it generally performs close to the optimal, demonstrating robustness and reliability.

  We further evaluated the effectiveness of \gopt's cardinality estimation method by comparing the performance of \gopt-plans utilizing \gloguequery, which incorporates high-order statistics, against those relying solely on low-order statistics.
  The results in \reffig{micro_exp_cardinality} show that, for 3 out of 8 queries, the plans utilizing high-order statistics achieved better performance, with an average speedup of $13.1\times$ across these three queries. High-order statistics were particularly effective for complex patterns such as $QC_{4a}$, where the plan utilizing high-order statistics was $31.9 \times$ faster.

  \stitle{{Optimizing Gremlin Queries.}}
  {We wrote the queries in Gremlin, enabling \gs~to generate GS-plans using its native rule-based optimizer and \gopt-plans after integrating \gopt. %By activating all the optimization techniques in \gopt,
  We compared the execution performance of GS-plans versus \gopt-plans on \gs,
  focusing on $QR$ and $QC$ queries, excluding $QT$ queries due to \gs's lack of type inference.
  As shown in \reffig{micro_exp_e2e}, for $QR$ queries testing heuristic rules, \goptplan~outperformed \gsplan~on most queries (except for $QR_5$ and $QR_6$, which test \joinelimrule, also supported by \gs's native optimizer), with an average speedup of $13.3\times$, even achieving significant improvements in $QR_2$ where \gsplan~runs \ot. This demonstrates the effectiveness of \gopt's new heuristic rules.
  For $QC$ queries, \gsplan~adhered to user-specified orders, most running \ot, whereas \goptplan, utilizing cost-based optimization techniques with improved cardinality and cost estimation, produced more efficient search orders and achieved an average speedup of $243.4\times$. Since GS-plans frequently result in suboptimal query plans, we excluded them from further comparisons in the subsequent experiments.
  These results demonstrate that integrating \gopt~ into \gs~ can immediately enhance its performance for executing Gremlin queries.
  Moreover, \gopt~ extends \gs~ by enabling support for Cypher queries.
  }

  \begin{figure}[t!]
    \begin{center}
    \setlength{\subfigcapskip}{-6bp}
    \subfigure[Time Cost of LDBC Queries on Neo4j]{\label{fig:neo4j_all}
    {\includegraphics[height=2.8cm, width=8cm]{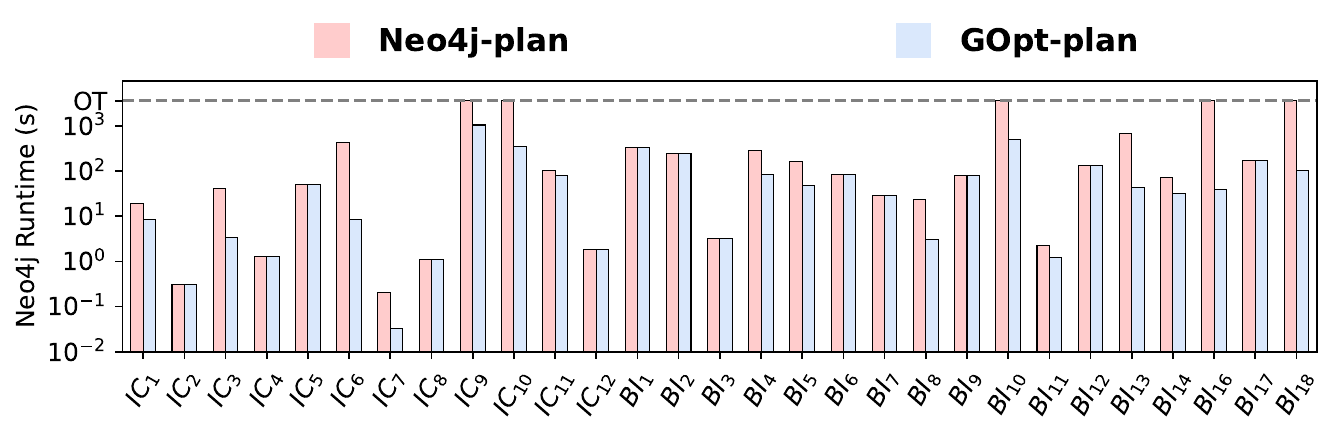}}}
    \setlength{\subfigtopskip}{-0.2cm}
    \setlength{\subfigcapskip}{-6bp}
    \subfigure[Time Cost of LDBC Queries on \gs]{\label{fig:gs_all}
    {\includegraphics[height=2.5cm, width=8cm]{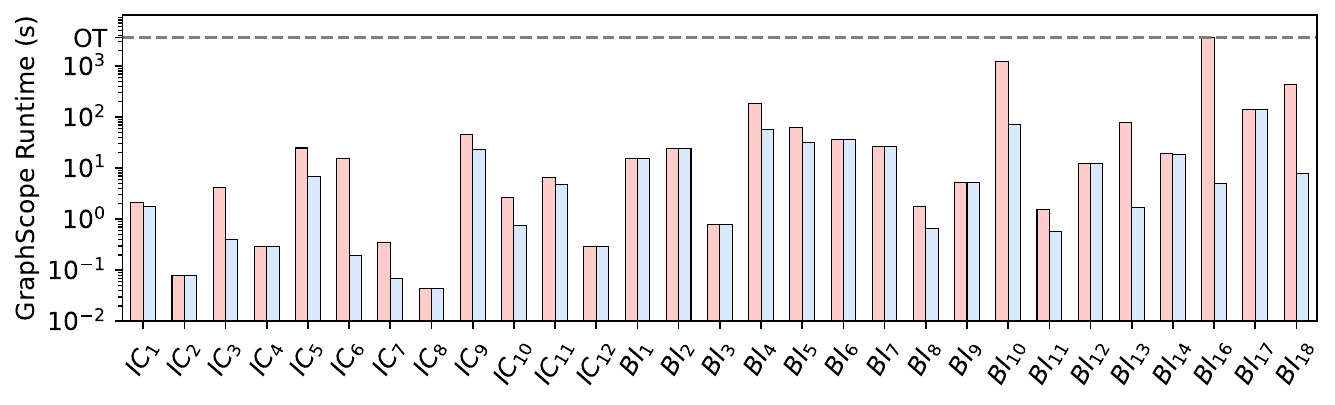}}}
    \caption{Results of Comprehensive Experiments.}
    \vspace{-2mm}
    \label{fig:comprehensive_exp}
    \end{center}
  \end{figure}

\vspace*{-2ex}
\subsection{{Comprehensive Experiments}}
\label{sec:comprehensive_exp}
{
  In this subsection, we compare \gopt~with Neo4j's \cypherplanner~by testing \gopt-plans and Neo4j-plans on both Neo4j and \gs.
  This comparison highlights \gopt's advanced optimization techniques and its ability to utilize backend-specific operators for superior performance.
  Experiments were conducted on $G_{100}$ using a single machine with 32 threads, as Neo4j supports only single-machine deployment.
  For fairness, we explicitly specify all type constraints in the queries, given Neo4j's schema-loose design and lack of type inference support in its optimizer.
}

  \stitle{\neoplan~v.s. \goptplan~on Neo4j.}
  {
    We began by comparing the performance of executing the \neoplan~and \goptplan~on Neo4j.
    The results in \reffig{neo4j_all} reveal that \goptplan~outperforms \neoplan~in $16$ out of $29$ queries, with an average speedup of $15.8\times$ across these queries, and $9.2\times$ overall.
    This enhancement is primarily due to \gopt's advanced optimization techniques.
    In $IC_{6}$, \gopt~achieves the most significant improvement by $48.6\times$ by employing a hybrid-join strategy, whereas Neo4j relies on multiple \physicalexpand, causing excessive intermediate results.
    In $IC_{10}$, $BI_{10}$ and $BI_{18}$, where \neoplan~runs \ot, \gsplan~optimizes  the search order with more precise cost estimation, reducing intermediate results and enhancing performance.
    Additionally, in $IC_{9}$ and $BI_{13}$, \gsplan~utilizes the \kw{AggregatePushDown} rule from Calcite, which pushes aggregation operations earlier to reduce intermediate results, a feature not supported by \cypherplanner.
    Due to the space limit, we will provide a detailed technical report to explain the performance gains of \gopt~over \cypherplanner~for each query.
    Overall, these experiments highlight the efficacy of \gopt's optimization techniques in improving query processing in Neo4j.
  }

  \stitle{\neoplan~v.s. \goptplan~on \gs.}
  {
    We further compared the performance of the two plans on \gs. Note that we manually translated Neo4j-plans into \gs-compatible versions to run on \gs.
    As shown in \reffig{gs_all}, \goptplan~outperformed \neoplan~in $17$ out of $29$ queries, with an average speedup of $56.2\times$ and $33.4\times$ overall, the most significant being $78.7\times$ in $IC_6$.
    This improvement is more pronounced than on Neo4j, as \gopt~enables the registration of optimized physical operators like \expandintersect~in \gs~via \physicalbuilder.
    For example, in the cyclic pattern query $IC_5$, both plans use \expandinto~on Neo4j and show similar performance.
    However, on \gs, \goptplan~selects the more efficient \expandintersect, achieving a $3.6\times$ speedup, whereas \neoplan~sticks with \expandinto, executed as a \physicalexpand~followed by \physicalselect.
    Other optimized queries also show improved performance on \gs~for similar reasons, additionally benefiting from the efficient \expandintersect.
    Overall, these results underscore the effectiveness of \gopt's optimization techniques in enhancing query processing on \gs.
  }

  \begin{figure}[t!]
    \begin{center}
    \setlength{\subfigcapskip}{-6bp}
    \subfigure[Data Scale of $IC$ queries on \gs]{\label{fig:data_scale_ic}
    \vspace*{-2ex}
      {\includegraphics[height=2.8cm, width=8cm]{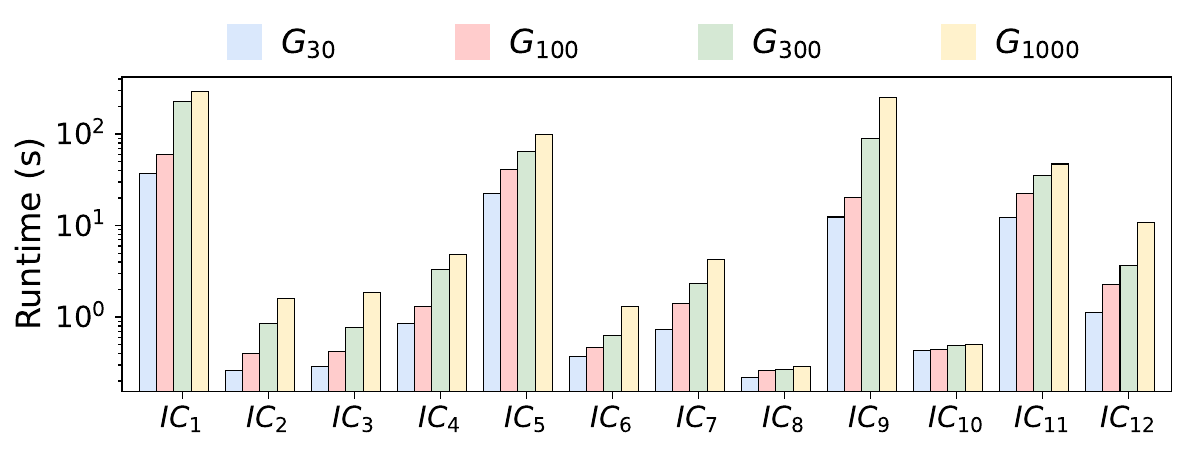}}}
      \setlength{\subfigtopskip}{-0.1cm}
    \setlength{\subfigcapskip}{-6bp}
    \subfigure[Data Scale of $BI$ queries on \gs]{\label{fig:data_scale_bi}
    \vspace*{-2ex}
      {\includegraphics[height=2.8cm, width=8cm]{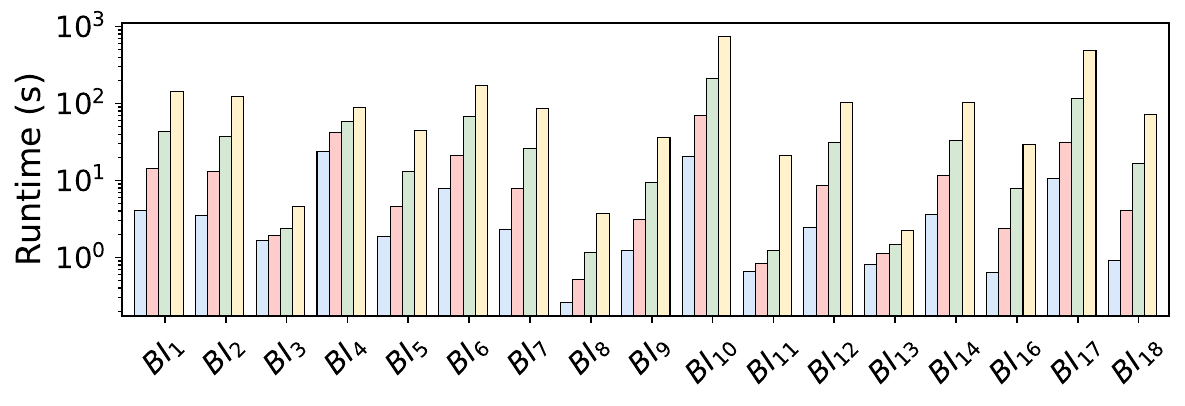}}}
    \vspace{-2mm}
    \caption{Results of Data Scale Experiments.}
    \vspace{-2mm}
    \label{fig:large_scale_exp}
    \end{center}
  \end{figure}

  \begin{figure}[t]
    \centering
    \includegraphics[height=2.8cm, width=8cm]{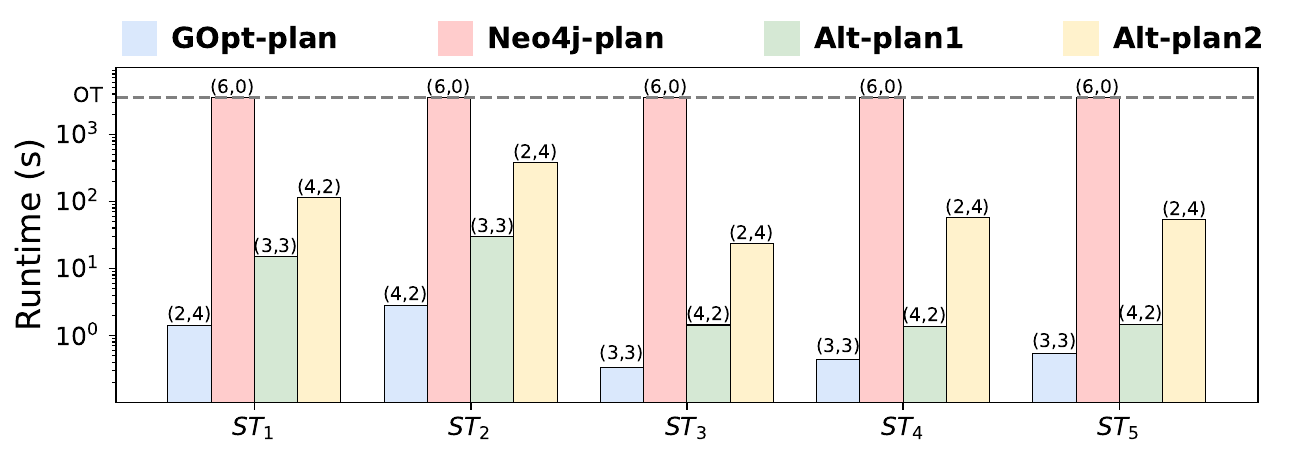}
    \caption{Performance of S-T Paths.}
    \label{fig:st_paths}
  \end{figure}

  \subsection{Data Scale Experiments}
  {
We also tested the scalability of \gs~integrated with \gopt, which supports distributed execution via \gaia, on a cluster of 16 machines, each equipped with 2 threads.
We conducted $IC$ and $BI$ queries across various dataset sizes. For $IC$ queries, we randomly selected $8$ parameter values per query, aggregating their runtimes to obtain a representative query runtime and mitigate bias from short execution times.
\reffig{data_scale_ic} shows the scalability results for $IC$ queries. When the graph size increases by $30\times$, from $G_{30}$ to $G_{1000}$, the runtime for the most impacted query ($IC_{9}$) increases by $20.0\times$, while the average performance degradation is only $6.3\times$, which is impressive given the substantial data scale.
\reffig{data_scale_bi} presents the results for $BI$ queries, demonstrating effective scalability with the growth of graph data.
As the graph size increases from $G_{30}$ to $G_{1000}$, the average performance degradation is $30.4\times$. This is expected, as $BI$ queries are typically more complex, causing intermediate results to grow steadily with the graph size.
These findings highlight the importance of effective query optimization, particularly the CBO, in preventing poor search orders that could generate intermediate results exponentially proportional to the graph size.
}

\subsection{Case Study}
\label{sec:case_studies}
%\stitle{Money Mule Detection.}
\gopt~with \gs~ backend is widely deployed in production at Alibaba.
This case study focuses on its application in fraud detection. Fraudsters transfer funds through multiple intermediaries before another fraudster withdraws the money.
%Given two fraudster set $S_1$ and $S_2$ and the hop number $k$, we aim to find all the money transformation paths between the fraudsters in $S_1$ and $S_2$ with the specified hop number.
This problem can be described as an s-t path in \patrel, with the Cypher query as follows:
% This problem can be formulated as a s-t path in \patrel, where the query in Cypher is as follows:
% \code{MATCH (p\_1:PERSON)-[p:*6]-(p\_2:PERSON) WHERE p\_1.id  and p\_2.id  RETURN p}
\begin{lstlisting}
  MATCH (p1:PERSON)-[p:*$k]->(p2:PERSON)
  WHERE p1.id IN $S1 and p2.id IN $S2
  RETURN p
\end{lstlisting}
Finding such paths in large graphs is challenging due to the explosion of intermediate results.
Two strategies are common:
(1) single-direction expansion, expanding $k$ hops from $S_1$ and filtering ends in $S_2$;
(2) bidirectional search, starting from $S_1$ and $S_2$ simultaneously, joining sub-paths when they meet in the middle.
% Two common strategies are:
% (1) single-direction expansion: start from vertices in $S_1$, expand $k$ hops, then filter to ensure the end vertices are in $S_2$;
% and (2) bidirectional search: Start traversal from $S_1$ and $S_2$ simultaneously and join sub-paths when they meet in the middle.
Bidirectional search is generally more efficient, but is the middle vertex in the path always the best choice for the join?
Our case study shows that it is not always true.

We conducted experiments on a real-world graph with $3.6$ billion vertices and $21.8$ billion edges.
We set $k=6$ and created five queries $ST_{1\ldots 5}$, each with distinct pairs of $(S_1,S_2)$ values randomly obtained from real applications.
For each query, {we generated \neoplan~and two alternatives that have different search orders to compare with \goptplan.}
The results, shown in \reffig{st_paths}, indicate the path join positions above each bar. For example, $(2,4)$ means a 2-hop sub-path from $S_1$ joins with a 4-hop sub-path from $S_2$, while
{all Neo4j-plans execute single-direction expansion from $S_1$ to $S_2$}.
\gopt-plans outperform all alternatives from $3\times$ to $20\times$, whereas all single-direction plans run \ot.
In \gopt-plans, the join position does not always lie at the center, as seen in $ST_1$ and $ST_2$, due to the varying number of source vertices in $S_1$ and $S_2$ in real queries.
By properly configuring the costs for scans (now they are the number of vertices in the source sets), \gopt's CBO
can automatically determine the optimal join positions.
%\gopt's CBO searches for the optimal plan by enumerating all possible join positions and selecting the one with the lowest estimated cost.
This demonstrate the effectiveness of \gopt~in optimizing \patrels~in real-world applications.

\eat{
\stitle{LDBC Interactive Complex Queries.}
\begin{figure}[t]
  \centering
  \includegraphics[width=\linewidth]{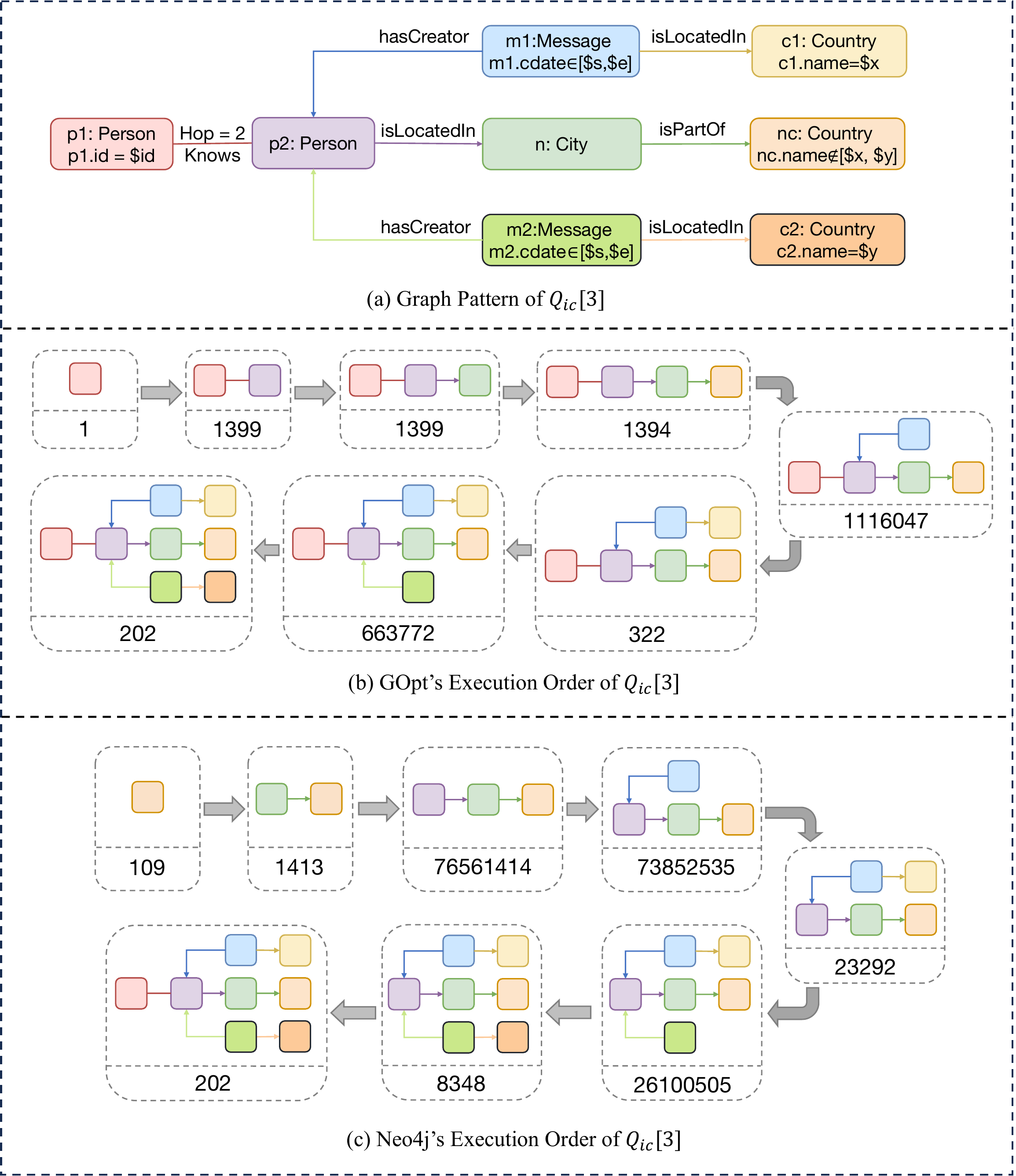}
  \caption{Execution Plans for $IC_3$}
  %\vspace{1mm}
  \label{fig:ldbc3_plans}
  \vspace*{-2mm}
\end{figure}
During our tests with the LDBC queries, \gopt~was observed to produce optimized plans that match the quality of manually optimized ones from prior research \cite{qian2021gaia}.
The case study, focusing on $IC_3$ presented in \reffig{ldbc3_plans}(a), offers an in-depth comparison between the plans optimized by \gopt~and Neo4j, depicted in \reffig{ldbc3_plans}(b) and \reffig{ldbc3_plans}(c), respectively.
In the optimized plans, the search order of each vertex is marked.
For example, \gopt~starts the searching from \code{Person} $p_1$, while Neo4j starts from \code{Country} $nc$.
%We assessed both plans' performance, measuring execution time and the number of intermediate results, and present the results in \reftab{case2}.
%\gopt's plan proved to be around 26 times faster than Neo4j's, with only 1\% of the intermediate results produced.
\eat{
\begin{table}
  \centering
  \vspace*{-0.5em}
  \caption{Comparison of $IC_3$ Physical Plans}
  \vspace*{-0.5em}
  \label{tab:case2}
  \begin{tabular}{ l | r | r }
  \hline
    \bf{Plan} & \bf{Avg. Runtime (s)} & \bf{Intermediate Result Num.} \\
    \hline
    \gopt's Plan & 6.085 & 1,784,334\\
    \hline
    Neo4j's Plan & 156.845 & 176,547,616 \\
    \hline
    \end{tabular}
\end{table}
}
Furthermore, we illustrate the number of intermediate results produced by each plan in \reffig{ldbc3_plans}, highlighting that \gopt's plan generates merely $1\%$ of the intermediate results compared to Neo4j's, while being around $26\times$ faster as verified in the previous small-scale experiments.
The main reason for the performance gap is that Neo4j's plan results in an explosion of intermediate results when expanding from \code{Person} $p_2$ to \code{Message} $m_1$ and $m_2$. In contrast, \gopt~expands $p_2$ from a user-specified starting point $p_1$, thereby limiting the number of potential matches for $p_2$ and effectively minimizing the matching numbers for subsequent expansions to \code{Message} $m_1$ and $m_2$. This case study underscores \gopt's efficacy in optimizing the execution of complex queries.
%This efficiency is mainly because Neo4j's plan generates a large number of intermediate results early in the process, while \gopt~strategically starts from a specific point, reducing early results and enhancing overall performance. This highlights \gopt's effectiveness in executing complex queries.
}

\vspace*{-1mm}
\section{Conclusion}
\label{sec:con}
\vspace*{-1mm}

In this paper, we present \gopt, a modular, graph-native query optimization framework for \cgps, which combine patterns and relational operations.
\cgps~can be expressed in multiple graph query languages and executed on various backends, necessitating a unified optimization approach.
\gopt~addresses this need by employing a unified \ir, and using a \graphirbuilder~to translate queries from diverse languages into this \ir, facilitating advanced optimization techniques.
To optimize \cgps, \gopt~utilizes comprehensive heuristic rules to refine the interactions between \bgps~and relational operations. 
It also includes an automatic type inference algorithm to uncover implicit type constraints in query patterns, and employs CBO techniques for further refinement. 
During the CBO process, \gopt~offers a \physicalbuilder, allowing backends to register custom physical operators and cost models, enabling more accurate cost estimation.
Extensive experiments demonstrate that \gopt~significantly improves query performance compared to state-of-the-art systems in both synthetic benchmarks and real-world applications.
%Integrated in Alibaba's \gs~platform, \gopt~has significantly improved query performance, demonstrating its valuable for business-critical applications, showcasing the practical and impactful benefits of \gopt~in a real-world production environment.

% \begin{acks}
%  This work was supported by the [...] Research Fund of [...] (Number [...]). Additional funding was provided by [...] and [...]. We also thank [...] for contributing [...].
% \end{acks}
%\clearpage
\bibliographystyle{ACM-Reference-Format}
\bibliography{main}
% \begin{appendix}
%     \input{sec-appendix}
% \end{appendix}
\end{document}